\pdfoutput=1
\documentclass[structabstract]{aa}  
\usepackage[english]{babel}  
\usepackage{graphicx}  
\usepackage[utf8]{inputenc}  
\usepackage{microtype}  
\usepackage{booktabs}  
\usepackage{multirow}
\usepackage{rotating}  
\usepackage[squaren]{SIunits}
\usepackage{natbib}
\usepackage{amsmath}
\usepackage{amssymb}
\usepackage{txfonts}
\usepackage{hyperref}
\usepackage{subfigure}

\newcommand{\tex}{T_\mathrm{{ex}}}
\newcommand{\tmb}{T_\mathrm{{MB}}}
\newcommand{\tbg}{T_\mathrm{{BG}}}

\newcommand{\tk}{T_\mathrm{{K}}}
\newcommand{\tb}{T_\mathrm{{B}}}
\newcommand{\tp}[2]{T_\mathrm{{#1}}^\mathrm{{#2}}}
\newcommand{\co}{C$^{18}$O}
\newcommand{\metac}{CH$_3$CCH}
\newcommand{\formt}{H$_2$CO(2$_{1,2}$-1$_{1,1}$)}

\newcommand{\pot}[1]{10^{#1}}
\newcommand{\solm}{\mathrm{M}_\odot}
\newcommand{\expo}[1]{\mathrm{e^{#1}}}
\newcommand{\vlsr}{V_\mathrm{{LSR}}}
\newcommand{\erg}{\mathrm{erg}}
\newcommand{\cm}{\centi \metre}

\newcommand{\eqn}[1]{Equation~(\ref{#1})}

\newcommand{\sbj}{\mathrm{J,J-1}}

\newcommand{\pc}{\mathrm{pc}}
\newcommand{\hii}{H\textsc{ii}}
\newcommand{\msun}{\mathrm{M_\odot}} 
\newcommand{\jy}{\mathrm{Jy}}
\newcommand{\yr}{\mathrm{yr}} 
\newcommand{\mg}{\mathrm{mag}}

\newcommand{\kms}{\kilo\metre\per\second}
\newcommand{\kel}{\usk\kelvin}
\newcommand{\mum}{\usk\micro\metre}
\newcommand{\asec}{$^{\prime\prime}$}

\bibpunct{(}{)}{;}{a}{}{,}

\title{Molecular clouds under the influence of massive stars in the Galactic \hii\ region G353.2+0.9 \thanks{This work is partially based on data taken at ESO (project 63.I-0189).}}
\institute{INAF Istituto di Radioastronomia, CNR, Via Gobetti 101, 40129, Bologna, Italy \label{ira}
\and Dipartimento di Astronomia, Universit\`{a} di Bologna, Via Ranzani 1, 40127, Bologna, Italy \label{unibo}
\and INAF Osservatorio Astrofisico di Arcetri, Largo E. Fermi 5, 50125, Firenze, Italy \label{arcetri}
\and Im Acker 21b, 56072 Koblenz, Germany \label{tief}}
\author{A. Giannetti \inst{\ref{ira}, \ref{unibo}}
\and J. Brand \inst{\ref{ira}} 
\and F. Massi \inst{\ref{arcetri}} 
\and A. Tieftrunk \inst{\ref{tief}}
\and M. T. Beltr\'{a}n\inst{\ref{arcetri}}} 

\abstract{}{We investigate the structure of the Galactic \hii~region G353.2+0.9, by analyzing (sub-)mm molecular-line and -continuum observations. This region is excited by the massive open cluster Pismis-24. We study the detailed morphology, distribution, and physical parameters (column and volume densities, masses, temperatures and opacities) of the molecular gas and dust. We are also interested in the variation in these parameters across the photon-dominated region.}
{We observed various molecules and transitions to derive the physical properties of the molecular gas through line ratios, and both LTE and non-LTE analyses. The physical properties of the gas were derived with a Bayesian approach for the non-LTE analysis. Based on the continuum data at $870\mum$, we derived the column density of molecular hydrogen from the surface brightness and thus molecular abundances from the molecular column densities. We determined the mass of the dust from the integrated flux. We also carried out the simplest possible analysis to identify the clump candidates for gravitational instability, determining their virial parameter $\alpha$.}
{The total mass of the gas in the region is $\sim 2000\usk\msun$, while that of the dust is $\sim21\usk\msun$. The presence of a velocity gradient in the region, with clumps with redder $\vlsr$ nearer Pis-24 suggests that the expansion of the ionized gas is pushing the molecular gas away from the observer.
We unambiguously identify the ionization front in G353.2+0.9, at the location of which we detect an increase in gas density and temperature. Its location and position angle is consistent with Pis-24 being the main ionization source. Almost no molecular gas is found south of the ionization front, at the location of the intense, elongated continuum and atomic-line emission, strengthening the hypothesis that Pis-24 is associated with G353.2+0.9.  
We find at least 14 clumps at different positions and LSR velocities, and we determine their physical conditions. The typical excitation temperatures are in the range of about $10-25\usk\kelvin$, while H$_2$ column densities are in the range $\sim \pot{20}-\pot{23}\usk\cm^{-2}$. 
From the non-LTE analysis, we derive kinetic temperatures in the ranges $11-45\kel$ (CS) and $20-45\kel$ (CN).  
The H$_2$ number density is typically around $\sim\pot{5}\usk\cm^{-3}$ from CS and $\mathrm{few}\times\pot{5}\usk\cm^{-3}$ from CN, with maxima above $\pot{6}\usk\cm^{-3}$. The abundances of the molecules observed are found to vary across the region, and appear to be higher in regions further away from the ionization front.
}
{}

\offprints{A. Giannetti (agianne@ira.inaf.it)}                                                                                              

\begin{document}  
\maketitle

\section{Introduction}
\begin{figure}[tb]
 \centering
 \includegraphics[angle=-90,width=\columnwidth]{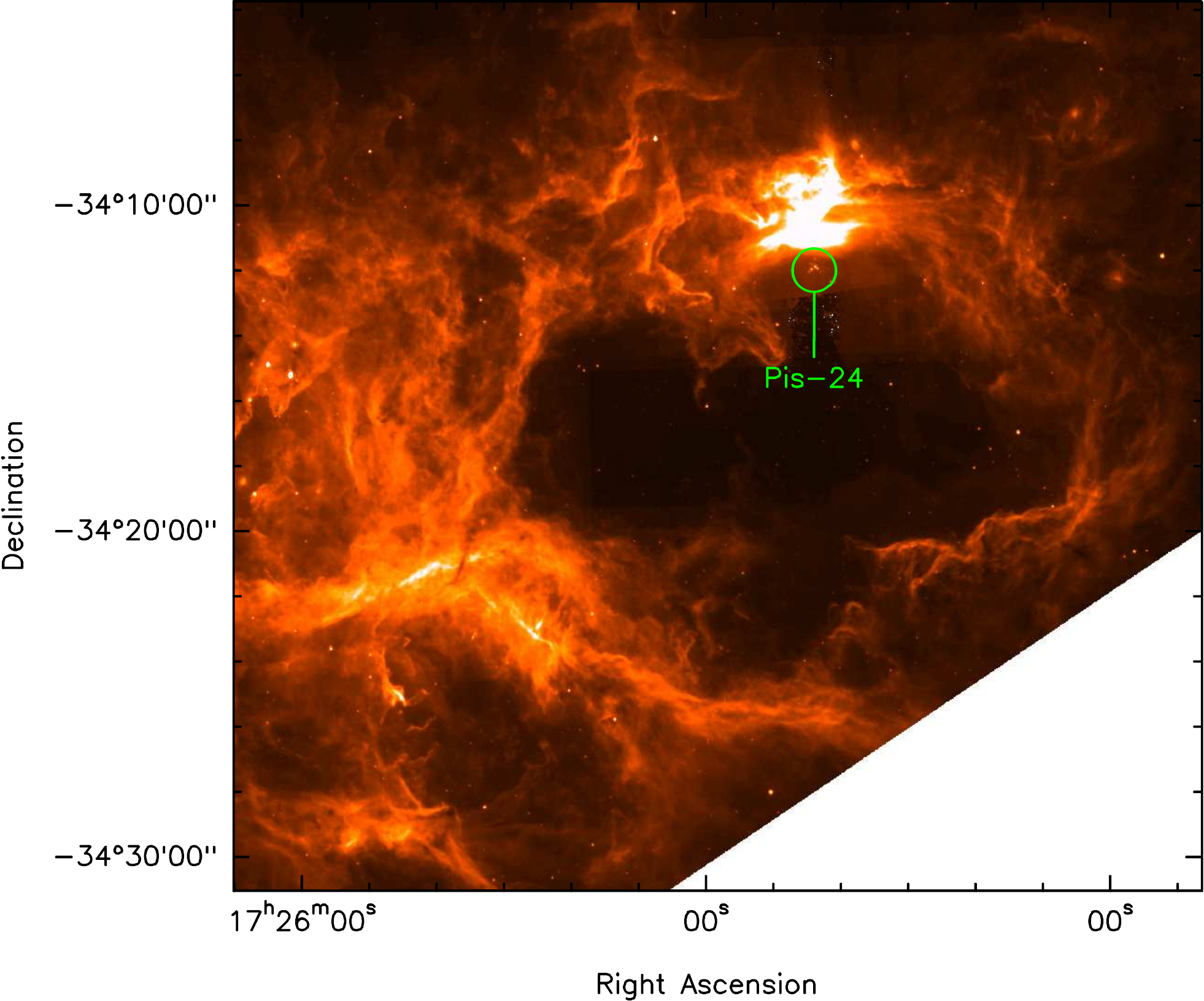}
 \caption{The $8 \usk \micro \meter$ emission image of NGC 6357, taken from the GLIMPSE survey (\url{http://www.astro.wisc.edu/sirtf/}, \citealt{Benjaminetal03}). The cavity is clearly visible at the centre of the image. G353.2+0.9 is the bright region at its northern border. The location of Pis-24 is also shown in the figure. The coordinates are referred to the epoch J2000.}
 \label{fig:GLI8.0mu}
\end{figure}

NGC 6357 is a complex of H{\sc ii} regions and molecular clouds that form a very active star-forming region in the Sagittarius spiral arm. Optical, radio, and infrared (IR) images of NGC 6357 confirm that it contains a number of distinct H{\sc ii} regions in different stages of evolution \citep[e.g.,][]{Fellietal90, Massietal97}. Figure~\ref{fig:GLI8.0mu} shows a large cavity or a collection of smaller, connected cavities in the region, delineated by ionized gas. Weak and diffuse H$\alpha$ emission permeates this feature. G353.2+0.9 is the bright emission region north of the cavity, seen in Fig.~\ref{fig:GLI8.0mu}. Just $55 \arcsecond$ south of G353.2+0.9, lies the massive open cluster Pismis 24 \citep[hereafter Pis-24; ][]{Pismis59}. This cluster is thought to be the main source of ionization of G353.2+0.9 \citep{Massietal97, Bohigasetal04}. It contains at least $\sim 20$ early-type (OB) stars, plus $24$ O-type candidates \citep{Wangetal07} and includes three stars that are amongst the brightest and bluest known in the Galaxy, of spectral types O3.5 III(f*), O3.5 If*, and O4 III(f+) \citep{Maiz-Apellanizetal07}. \citet{Masseyetal01} derived a distance of $2.56$ kpc and an age of $\sim1\usk\mega\yr$ for this cluster, and assuming that the molecular material is associated with Pis-24, we consider NGC6357 to be at the same distance. The large cavity is unlikely to have been formed by Pis-24, because of its clearly off-centre position. The morphology and the size of the cavity seem to suggest that it was shaped by the winds and/or supernova events of one or more clusters \citep{Wangetal07} situated in the proximity of the centres of the smaller bubble-like structures. 

\citet{Massietal97} performed a detailed study of the molecular emission associated with two of the \hii\ regions in NGC6357, to wit: G353.2+0.6, and G353.2+0.9. The latter region is the younger one and exhibits signs of the presence of recently formed massive stars [e.g., ultra-compact H{\sc ii} regions (UCH{\sc ii}), embedded sources with infrared (IR)-excess]. 

\begin{figure}[tb]
 \includegraphics[angle=-90,width=\columnwidth]{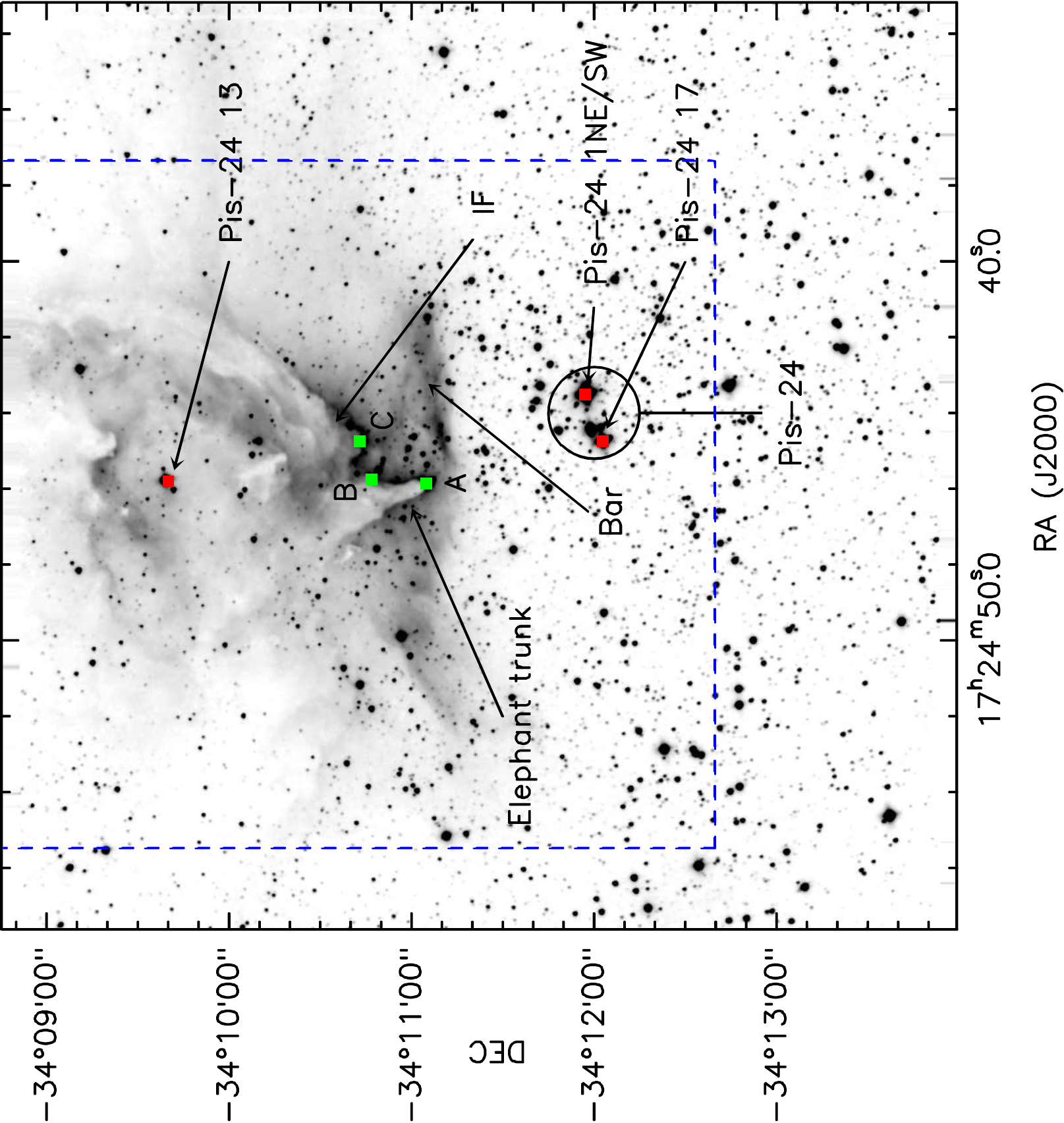}
\caption{$K_\mathrm{s}$-band image of G353.2+0.9 obtained with SofI (Massi et al., in prep.). The actual ionization front is indicated by ``IF'' (see text). The locations of the ``Bar'', Pis-24, and the elephant trunk are indicated. The red and green squares mark the location of some early-type stars in Pis-24 and the three UC\hii\ regions identified by \citep{Fellietal90}, respectively. The coordinates are referred to the epoch J2000.}
\label{fig:kbandif}
\end{figure}

Our study is focused on G353.2+0.9. \citet{Fellietal90} observed it with the VLA at $\lambda = 6$ cm with an HPBW of $3\farcs5$ (Fig.~7a of \citealt{Fellietal90}).
As for the H$\alpha$ emission, the high-resolution interferometric radio continuum observations reveal a very complex structure of the ionized gas, with a well-defined sharp boundary running east-west (the ``Bar'', in Figure~\ref{fig:kbandif}). The emission is characterized by a strong intensity gradient to the south, while showing a more gentle decrease to the north (Fig.~8 in \citealt{Fellietal90}).
\citet{Fellietal90} found three UCH{\sc ii} (A, B, and C in Fig.~\ref{fig:kbandif}).

The $K_\mathrm{s}$ band image in Fig.~\ref{fig:kbandif} shows that in the central part of the nebula there is an elephant trunk-like region of obscuration (clearly visible also in HST images; \citealt{HesterDesch05}), with an UC\hii\ region and IR source at its apex. This source shows a near-IR excess and X-ray emission: it was identified from HST observations to be in the evaporating gaseous globule (EGG) evolutionary phase \citep{Hesteretal96}, making it the first X-ray emitting EGG. This embedded object was classified as having a spectral type B0-B2 \citep{Wangetal07}. The elephant trunk points toward Pis-24 and is thought to be formed by the radiation and stellar winds from the OB stars in this cluster. The IR emission is brightest along the sides of the trunk and on the south-western side of G353.2+0.9, facing Pis-24. 

\citet{Massietal97} mapped G353.2+0.9 in CO(1-0) and $^{13}$CO(1-0). These data were complemented with observations of other molecules and transitions along strips in the north-south direction, to determine variations in physical parameters across the photon-dominated region (PDR). \citet{Massietal97} found that G353.2+0.9 is a face-on, blister-type H{\sc ii} region, with most of the molecular material behind the H{\sc ii} region and to the north of it.
\citet{Fellietal90} suggested that G353.2+0.9 is not associated with Pis-24, arguing that the southern sharp boundary (the ``Bar'') is produced by local ionization caused by embedded sources, that are also responsible for the radio emission of G353.2+0.9. They also concluded that the nebula is ionization-bounded to the south, implying that there are considerable quantities of molecular gas in the region south of the ionization front. Molecular-line observations do not however support this claim: \citet{Massietal97} found very little molecular emission at the location of the ``Bar''.

\citet{Bohigasetal04} found that this elongated structure has to have a considerable extent along the line-of-sight ($1-5 \usk \pc$, \citealt{Bohigasetal04}). While this dimension is comparable to the extent in R.A., it is much larger than the extent in DEC. This suggests that the ``Bar'' is a layer of ionized matter seen edge-on. It could be caused by the interaction of the photoionized photoevaporative flow with the free wind of the Pis-24 stars \citep{Healyetal04}. This implies that the molecular gas in the region has already been swept by the stellar winds, i.e. the region south of the ``Bar'' should be nearly devoid of molecular material.

The present study follows up on the work described in \citet{Massietal97}, which constitutes a first step in the study of the interface between the H{\sc ii} region and the molecular cloud.
Our aim is to clarify the morphology of the region, by observing optically thin molecular lines (e.g. C$^{18}$O), and to confirm the absence of molecular material south of the main ionization front, which is identified as IF in Fig.~\ref{fig:kbandif} (see Sect.~\ref{ssec:geomif}). This would strengthen the hypothesis of an association between G353.2+0.9 and Pis-24.
The physical conditions of the gas are derived by means of a non-LTE analysis for those molecules with several observed transitions or with hyperfine structure, while for the remaining molecules we assumed LTE (local thermodynamic equilibrium). The observation of the continuum at $870\usk\micro\metre$ allows us to infer the dust column density and mass, and thus to determine those of the gas, by assuming a gas-to-dust ratio. With the H$_2$ column densities determined in this way, we were able to calculate the abundance for the observed molecules.

\section{Observations and data reduction}

The molecular-line observations were carried out between Sept. 1 and 9, 1999, with the 15-$\metre$ Swedish-ESO Submillimeter Telescope (SEST; program 63.I-0189). This radio telescope was operational on La Silla in the period 1987-2004.
The telescope was equipped with SIS receivers and a high-resolution acousto-optical spectrometer with a total bandwidth of $86 \usk \mega \hertz$, and a frequency resolution of about $42 \usk \kilo \hertz$. The spectrometer was split into two parts, and we observed with two receivers simultaneously, one at a lower and the other at a higher frequency [e.g., \co(1-0) and \co(2-1)]. Detailed information about the molecular transitions observed are reported in Table~\ref{tab:transit}. The columns indicate the molecules and transitions observed, their rest frequency, resolution in frequency and velocity, beam FWHM, main beam efficiency, spacing between points in the maps, and typical $\tmb$ rms noise per channel of the spectra, respectively.

Most observations were made in frequency-switching mode, with a switch-interval in frequency sufficiently small to make the emission appear in both the signal and reference cycles, but large enough to avoid overlap between them. 

The SiO(5-4), CN(1-0), CN(2-1), and CH$_3$CCH(6-5) lines were observed with position switching. The emission of the last two molecules exhibits hyperfine structure, making it necessary to observe in position-switch mode. The telescope pointing was checked every three hours on the nearby SiO maser source AH Sco, and was found to be accurate to within $5 \arcsecond$. The same source was also used as the off-position. The calibration was made using the standard chopper-wheel method described in \citet{KutnerUlich81}. 
Our molecular-line maps are centred on $\alpha = 17^\mathrm{h} 24^\mathrm{m} 45\fs6$, $\delta = -34\degree 11' 20\farcs7$ (J2000), coinciding with the ``Bar''. The angular extent of the observed region is about $5'\times5'$ for CS(2-1) and (3-2), while it is $\sim 3'\times3'$ for the other molecules and transitions, with spacing between the raster point listed in Table~\ref{tab:transit} (Col. 7). 
The line intensities in this paper are expressed in terms of the main beam temperature, defined as $\tmb=\tp{A}{*}/\eta_\mathrm{MB}$. Data reduction and analysis for molecular-line data were performed with CLASS, part of the GILDAS (Grenoble Image and Line Data Analysis Software\footnote{\url{http://iram.fr/IRAMFR/GILDAS/}}) package.

The morphology and the distribution of the molecular gas can be investigated in greater detail by decomposing the emission profile into single Gaussian components at different $\vlsr$. 
Given the limited number of velocity components, we decided to decompose the emission profiles at every position by fitting different Gaussian curves, starting from optically thin transitions (e.g. \co). The characteristics of the components identified in this way were then used as a template for the decomposition of the  emission profiles of the other molecules and transitions.

CLASS offers the possibility to fit lines with hyperfine structure, such as those of CN, by specifying the relative intensity of the hyperfine components in the case of optically thin emission and assuming that their ratios have their LTE values. CLASS uses the optical depth as a free parameter of the fit, and gives it as output of the procedure. 

\medskip
\begin{table}[t]
 \centering 
\scriptsize
\caption{Molecular transitions observed.} 
\label{tab:transit} 
\medskip 
\begin{tabular}{cccccccc} 
\toprule 
Transition & $\nu$ & $\Delta\nu$ & $\Delta V$ & $\theta_\mathrm{MB}$    & $\eta_\mathrm{MB}$ & Sp.  & rms \\
                    & (GHz)     & (kHz) & {($10^{-2}$km/s)}       & $(\arcsecond)$               &                 & $(\arcsecond)$     & $(\kelvin)$ \\
\midrule
C$^{18}$O(1-0)              & $109.7822$  & $41.7$      & $11.4 $ & $47$ & $0.70$          & $25$          & $0.11$ \\
C$^{18}$O(2-1)              & $219.5603$  & $41.7$      & $5.70 $ & $24$ & $0.50$          & $25$          & $0.20$ \\
C$^{34}$S(2-1)              & $96.4130$   & $41.7$      & $13.0 $ & $54$ & $0.75$          & $25$          & $0.10$ \\
CS(2-1)                     & $97.9810$   & $41.7$      & $12.8 $ & $53$ & $0.75$          & $50$          & $0.10$ \\
CS(3-2)                     & $146.9691$  & $41.7$      & $8.51 $ & $35$ & $0.66$          & $50$          & $0.13$ \\
CS(5-4)                     & $244.9356$  & $41.7$      & $5.11 $ & $21$ & $0.50$          & $25$          & $0.25$ \\
H$_2$CO(2$_{1,2}$-1$_{1.1}$)& $140.8395$  & $41.7$      & $8.88 $ & $37$ & $0.66$          & $25$          & $0.30$ \\
CN(1-0)                     & $113.4910$  & $41.7$      & $11.0 $ & $46$ & $0.70$          & $25$          & $0.11$ \\
CN(2-1)                     & $226.8748$  & $41.7$      & $5.51 $ & $23$ & $0.50$          & $25$          & $0.23$ \\
CH$_3$CCH(6-5)              & $102.5401$  & $41.7$      & $11.2 $ & $51$ & $0.70$          & $...^{(1)}$   & $0.17$ \\
SiO(5-4)                    & $217.1049$  & $41.7$      & $5.76 $ & $24$ & $0.50$          & $...^{(1)}$   & $0.18$ \\
\bottomrule 
\end{tabular} 
\\
\medskip
\textsc{Note ---} $^{(1)}$ Observed at selected positions only.
\end{table}

We retrieved a map of G353.2+0.9 at $870\usk\micro\metre$ ($345\usk\giga\hertz$), taken with APEX from the ATLASGAL survey \citep{Schulleretal09}. The rms noise in the map is $\sim 100 \usk \milli \jy \per \mathrm{beam}$, determined in three regions free of emission around the \hii~region.
These data were analyzed with MOPSIC, the evolution of MOPSI (Map On-off Pointing Skydip Image), which was developed by R.~Zylka (Obs. de Grenoble).

\section{Results and discussion}

\subsection{Morphology}

\begin{figure*}[tbp]
 \centering
 \subfigure{\includegraphics[angle=-90, width=0.80\textwidth]{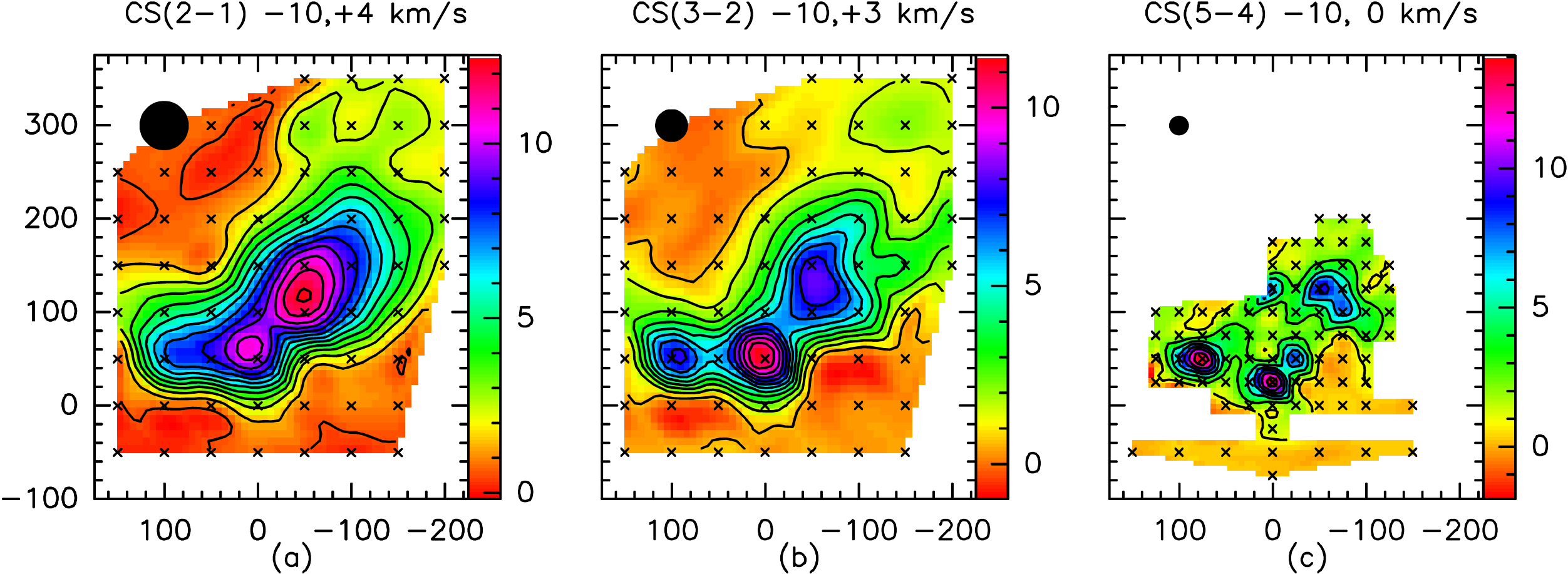}}\\
 \subfigure{\includegraphics[angle=-90, width=0.80\textwidth]{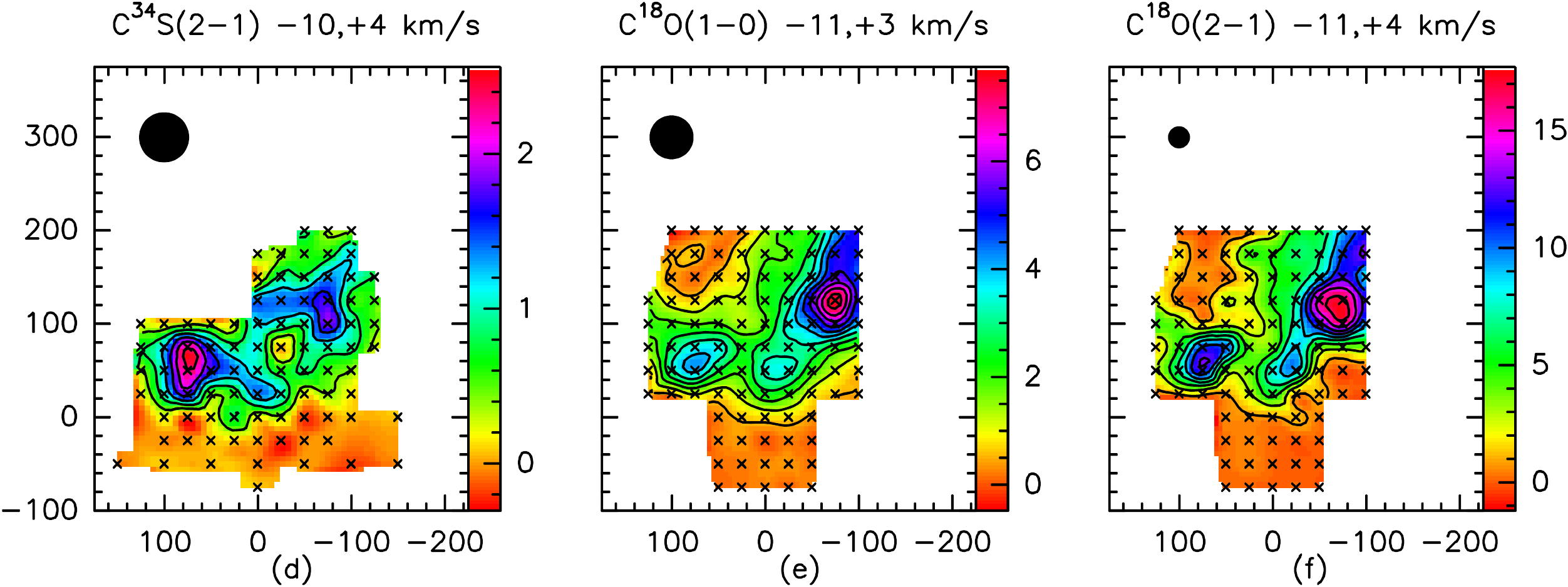}}\\
 \subfigure{\includegraphics[angle=-90, width=0.80\textwidth]{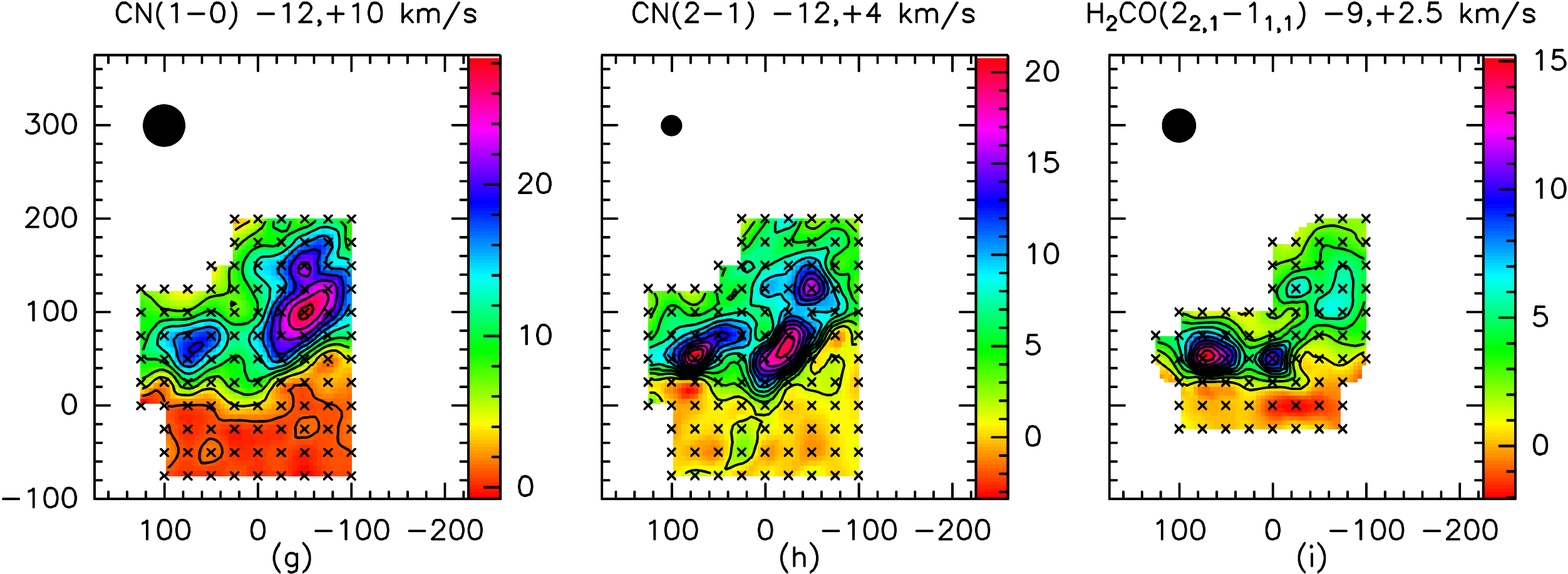}}\\
 \caption{Maps of the integrated emission $\int \tmb \mathrm{d}\upsilon$ of the different molecular species and transitions, within the whole velocity range of emission. The first contour is the $3\sigma$ level. Molecule, transition, and integration limits are indicated above each map. The beam size is indicated by the filled circle. The observed positions are marked with a cross. The contours levels (in units of $\kelvin \usk \kilo \metre \per \second$) are, respectively (lowest (step) highest): (a) $0.41 $ (1.0) $12.41 $, (b) $0.51 $ (1.0) $11.51 $, (c) $0.89 $ (2.0) $12.89 $, (d) $0.4 $ (0.4) $2.4 $, (e) $0.47 $ (0.7) $7.47 $, (f) $0.81 $ (2.0) $16.81 $, (g) $0.98 $ (3.0) $27.98 $, (h) $1.25 $ (2.0) $19.25 $, (i) $0.92 $ (1.5) $15.92 $. Coordinates are offsets (arcsec) with respect to $17^\mathrm{h} 24^\mathrm{m} 45\fs6$ $\delta = -34\degree 11' 20\farcs7$ (J2000).}
 \label{fig:intem}
\end{figure*}

Figure~\ref{fig:intem} shows the maps of integrated line-emission $\int \tmb \mathrm{d}\upsilon$ towards G353.2+0.9.
The molecular emission never extends significantly below $\Delta \delta = 0 \arcsecond$, confirming the lack of molecular material south of the ``Bar'' and ruling out the possibility that this feature is an ionization front proceeding southward.

\begin{figure*}[tbp]
 \centering
 \includegraphics[angle=-90, width=0.90\textwidth]{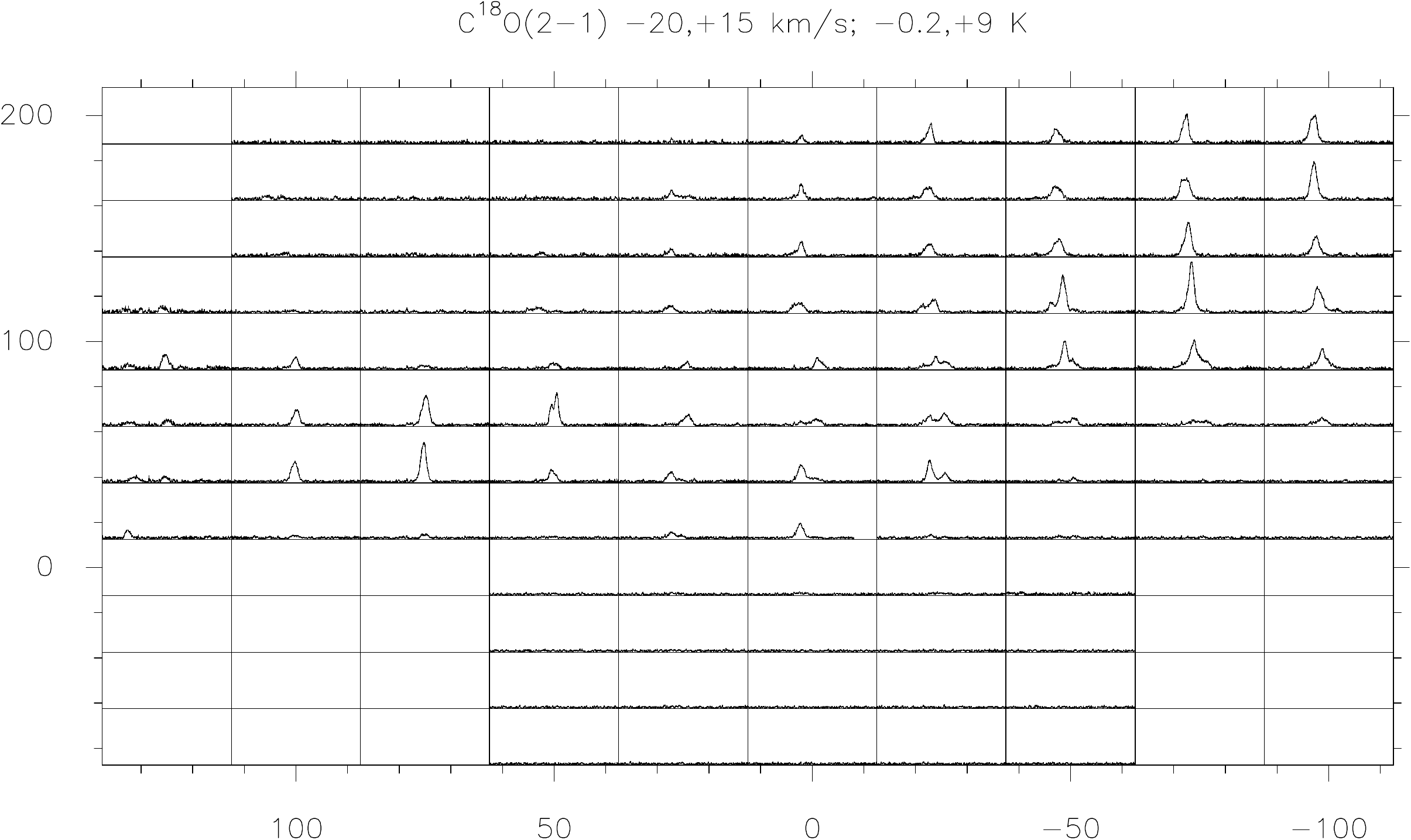}
 \caption{$\tmb$(\co(2-1)) spectral map. Each frame corresponds to a single pointing with an offset in $\alpha$ and $\delta$ indicated in the figure. The $x$ and $y$ axes of each spectrum range from $-20$ to $15\usk\kilo\metre\per\second$ and from $-0.2$ to $+9\usk\kelvin$, respectively.}
 \label{fig:specmap}
\end{figure*}

The molecular emission is concentrated between $-10$ and $+1 \usk \kilo \metre \per \second$ with, in some cases, strongly varying emission profiles between adjacent positions, as clearly visible in Fig.~\ref{fig:specmap}. 
In Fig.~\ref{fig:intem}, it is possible to identify many different clumps.
In higher-frequency transitions, such as CS(5-4) and CN(2-1), several clumps are resolved into two or more smaller clumps, or show an elongated appearance. 

\begin{figure*}[tbp]
 \centering
 \subfigure[\label{fig:c18o} \co(1-0)]{\includegraphics[angle=-90, width=0.75\textwidth]{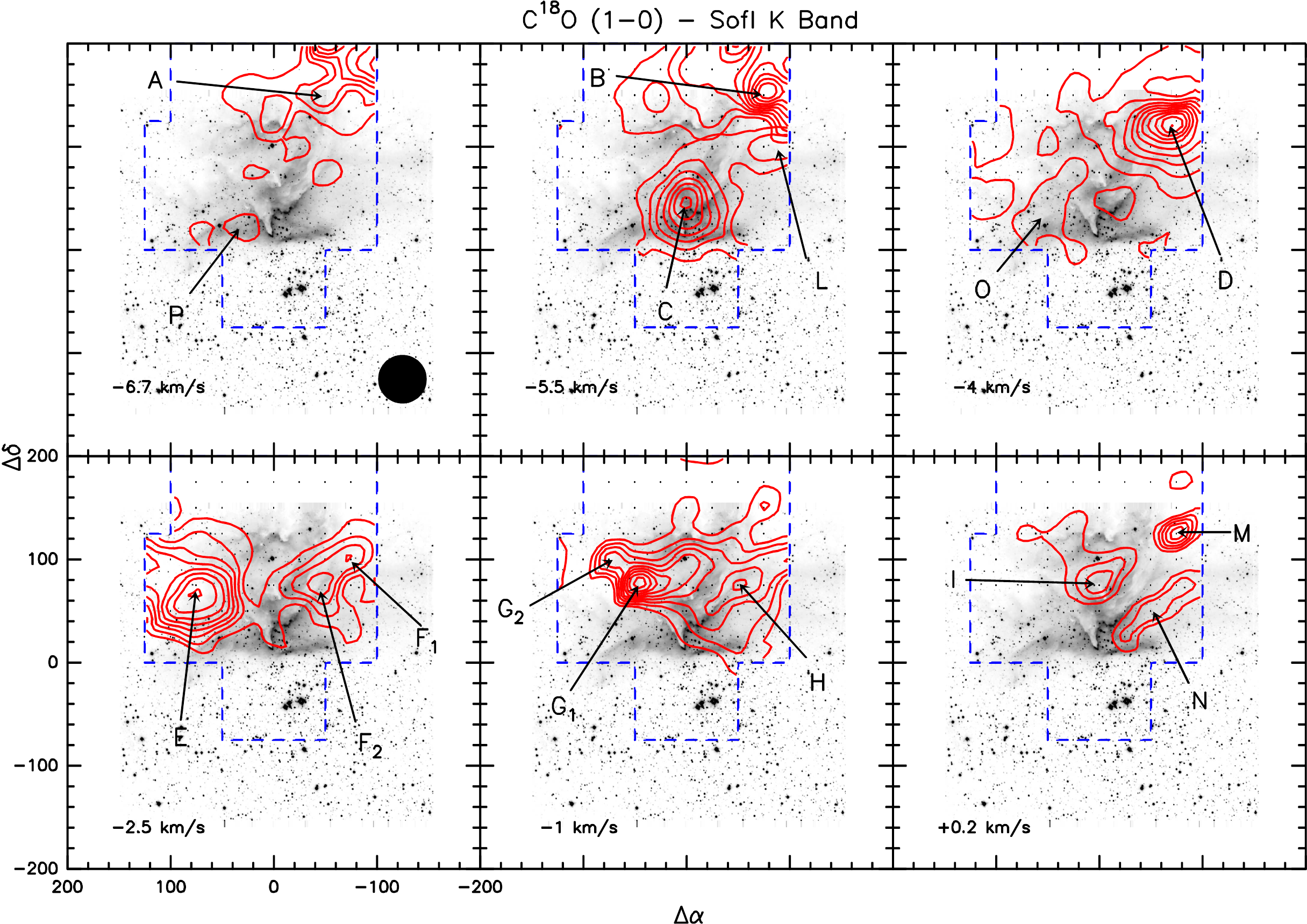}}
 \subfigure[\label{fig:cs} CS(5-4)]{\includegraphics[angle=-90, width=0.75\textwidth]{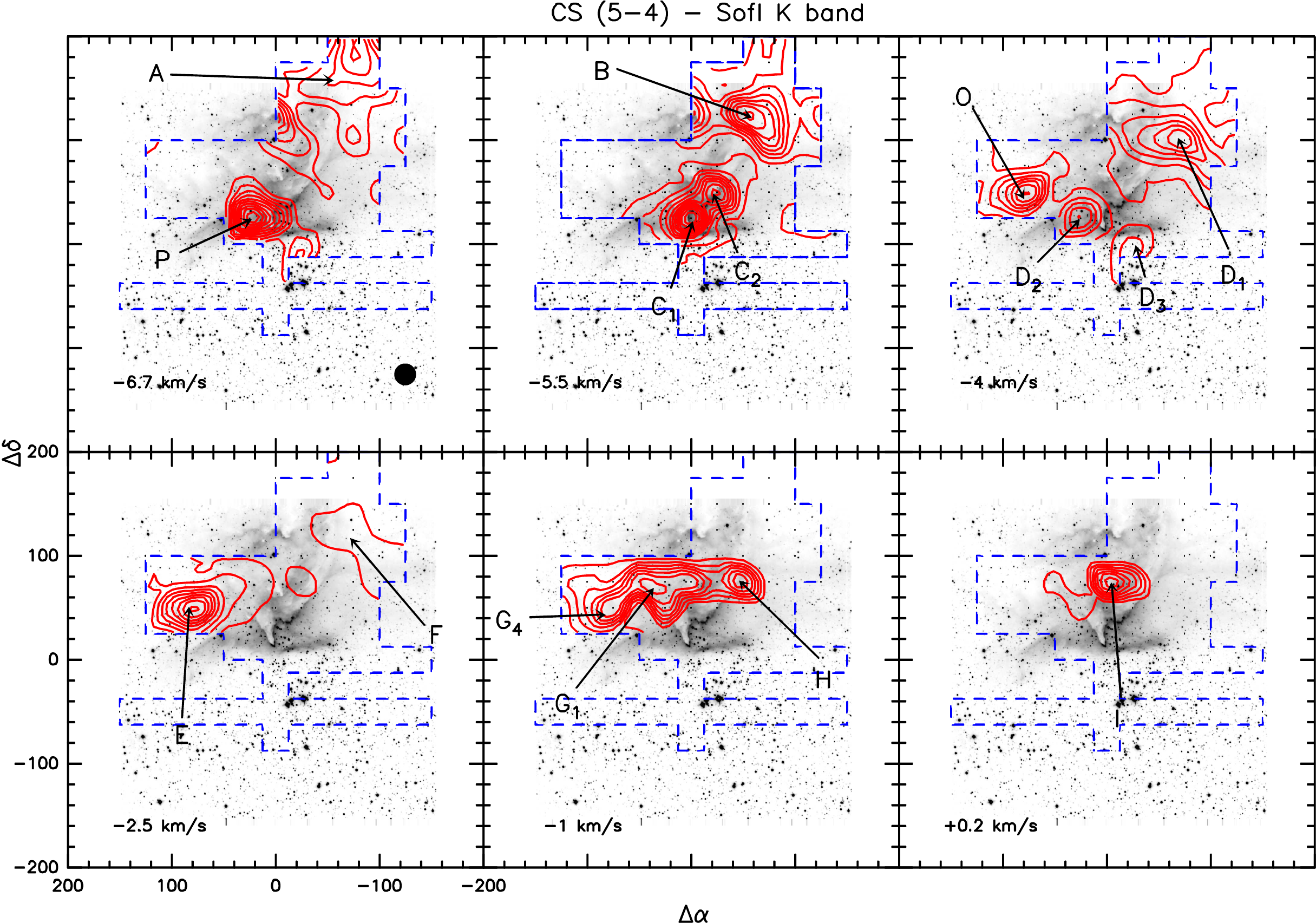}}
 \caption{\textbf{(a)} Integrated emission $\int \tmb \mathrm{d}\upsilon$ of \co(1-0) superimposed on the $K_\mathrm{s}$ image. Each plot corresponds to a different $\vlsr$. The $\vlsr$ of the clumps is indicated in the respective panel. The dashed (blue) lines indicate the observed area. The (red) contours show the integral under Gaussian components fitted to the line profiles. The beam of the transition is shown in the first panel. The first contour in each panel is the $3\sigma$ level. The integrated emission levels (in units of $\kelvin\usk\kilo\metre\per\second$) are (lowest (step) highest): ($-6.7\usk\kilo\metre\per\second$) 0.17 (0.3) 2.27; ($-5.5\usk\kilo\metre\per\second$, C) 0.19 (0.3) 1.99; ($-5.5\usk\kilo\metre\per\second$, B) 0.20 (0.5) 4.70; ($-4.0\usk\kilo\metre\per\second$) 0.18 (0.7) 6.48; ($-2.5\usk\kilo\metre\per\second$) 0.19 (0.3) 2.89; ($-1.0\usk\kilo\metre\per\second$) 0.18 (0.2) 1.78; ($+0.2\usk\kilo\metre\per\second$) 0.15 (0.1) 0.65.
 \textbf{(b)} As \textbf{(a)}, but for CS(5-4). The integrated emission levels (in units of $\kelvin\usk\kilo\metre\per\second$) are (lowest (step) highest): ($-6.7\usk\kilo\metre\per\second$) 0.19 (0.3) 2.59; ($-5.5\usk\kilo\metre\per\second$, C) 0.20 (1.0) 10.20; ($-5.5\usk\kilo\metre\per\second$, B) 0.19 (0.5) 3.69;  ($-4.0\usk\kilo\metre\per\second$) 0.19 (0.7) 4.39; ($-2.5\usk\kilo\metre\per\second$) 0.18 (1.0) 8.18; ($-1.0\usk\kilo\metre\per\second$) 0.17 (0.2) 1.37; ($+0.2\usk\kilo\metre\per\second$) 0.14 (0.1) 0.74. Note that names of the clumps do not correspond to those used in \citet{Massietal97}.}
 \label{fig:cs-c18o}
\end{figure*}

We found that the emission can be separated into 14 clumps in six velocity ranges (see Figure~\ref{fig:cs-c18o}), i.e. $-7.3 \lesssim \vlsr \lesssim -6.1$ (A and P), $-6.1 \lesssim \vlsr \lesssim -4.7$ (B, C and L), $-4.7 \lesssim \vlsr \lesssim -3.3$ (D and O), $-3.3 \lesssim \vlsr \lesssim -1.9$ (E and F), $-1.9 \lesssim \vlsr \lesssim -0.4$ (G and H), and $-0.4 \lesssim \vlsr \lesssim +0.8$ (I,M and N). The names of the clumps do not correspond to those used in \citet{Massietal97}, because of the higher spatial resolution in our present work and a different method of analysis (Gaussian decomposition versus integrated emission in velocity bins).

Figure~\ref{fig:cs-c18o} shows that the various clumps have slightly different position for different molecules and transitions, which could be the result of different excitation conditions, optical depth, chemical, or resolution effects. Furthermore, some clumps (such as P and O) are clearly visible only in high-density, high-frequency tracers. This is particularly clear in CS (5-4), which has the highest resolution. The emission of these tracers shows the presence of multiple high-density, small cores within a larger clump.

The clumps along the ionization front tend to have redder velocities than the others.
This is especially true for low-density tracers (cf. Fig.~\ref{fig:cs-c18o}). Clumps N, H, and F, and D, B, and A exemplify this behaviour. This can be understood by taking into account the radiative and mechanical action of the stars of Pis-24:
when neutral gas is exposed to the intense energetic radiation of an early-type star, it becomes rapidly ionized near the surface of the cloud. This gas is heated to $T\sim10000\usk\kelvin$ (a factor of $\sim 100$ with respect to cold, neutral gas in the cloud), consequently causing a comparable increase in pressure, thus leading to a rapid expansion. However, the expansion towards the neutral gas is stopped by the presence of the dense material of the cloud. In the opposite direction, the low-density ionized gas cannot halt the expansion of this overpressurized gas. The ionized material moves predominantly away from the cloud, having an equal and opposite effect on the cloud.

Therefore, the molecular emission indeed suggests that the ionized, overpressurized gas in G353.2+0.9, observed by \citet{Bohigasetal04}, is expanding, thus pushing the molecular material away from the observer.

\subsection{Temperatures}

The first method used to derive the excitation temperatures uses the ratio of main beam temperatures of the (2-1) to (1-0) transitions of \co. We first resampled the spectra to the same velocity resolution, and then convolved the \co(2-1) data with a Gaussian beam to match the spatial resolution of the (1-0) transition.

We calculated the ratio $R_1$ of $\tmb$(2-1) to $\tmb$(1-0) and determined $T_\mathrm{ex}$, assuming optically thin emission, from \citep{Levreault88}
\begin{equation} 
 R_1 = 4 \mathrm{e}^{-h \nu_{21} / k T_{\mathrm{ex},21}} = 4 \mathrm{e}^{-10.50 / T_{\mathrm{ex},21}},
 \label{eqn:r1s}
\end{equation}
where $T_{\mathrm{ex},21}$ is the excitation temperature of the \co(2-1) transition. Assuming LTE, $\tex$ is the same for all levels. 
Since both transitions were observed simultaneously, the systematic calibration uncertainties should not affect the temperature ratios, which are instead influenced by uncertainties in the Gaussian decomposition of the emission profile. Taking into account the uncertainties in the Gaussian fit performed with CLASS, we estimated that the uncertainties in $\tmb$ should not exceed $5-15 \%$, where both lines are detected above $3\sigma$. Assuming a typical uncertainty in $\tmb$ of $10\%$, the relative uncertainty in $\tex$ is $\sim 14\%$.
We found that $R_1$ is always between $\sim 2$ and $\sim 4$, indicative of optically thin emission throughout the region. The $\tex$(\co) (derived from Eq.~\ref{eqn:r1s}) is fairly uniform, typically between $\sim15 \usk \kelvin$ and $\sim25 \usk \kelvin$ for all clumps. 
The higher values are found along the ionization front, while for clump C, which is associated with the elephant trunk, we have $\tex\sim20\kel$.

\begin{figure}[tbp]
 \includegraphics[angle=-90, width=\columnwidth]{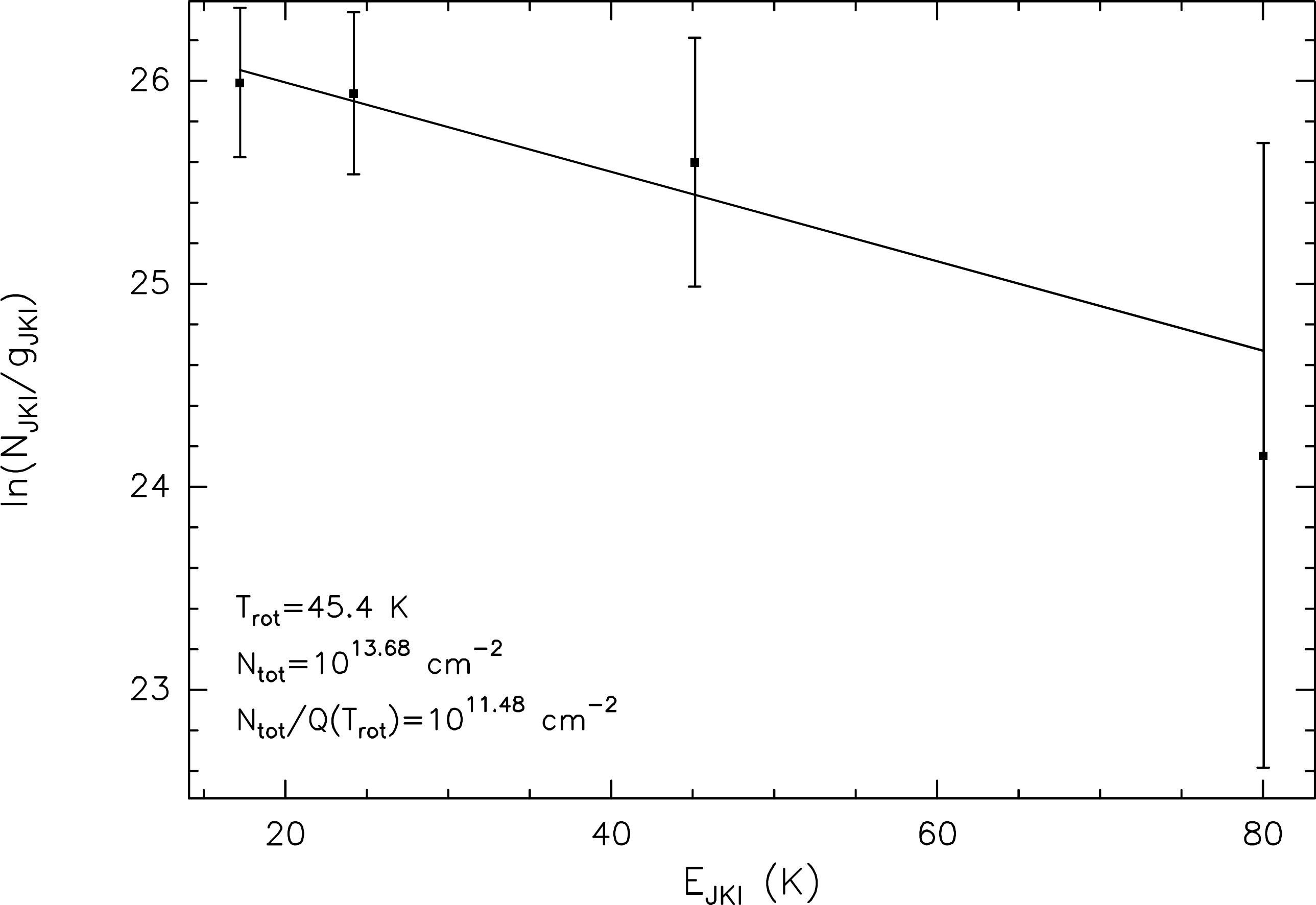}
 \caption{Boltzmann plot obtained from the \metac\ data at ($0\arcsecond$,$50\arcsecond$).}
 \label{fig:rotdia}
\end{figure}

Another temperature probe is methyl acetylene (\metac) \citep{Berginetal94}. This molecule has a high critical density, and its emission comes from high-density regions.
Our observations are limited to four positions: ($0\arcsecond$,$50\arcsecond$) (clump C), ($75\arcsecond$,$75\arcsecond$) (clump E), ($-75\arcsecond$,$125\arcsecond$) (clump B), and ($0\arcsecond$,$-25\arcsecond$). We detected \metac~only at ($0\arcsecond$,$50\arcsecond$) and ($75\arcsecond$,$75\arcsecond$), while the detection at ($-75\arcsecond$,$125\arcsecond$) is uncertain. 
Only at the first position were four components of the K-ladder visible, giving a reliable temperature estimate. The Boltzmann plot analysis (Fig.~\ref{fig:rotdia}) gave a $T_\mathrm{ex}\sim 45\usk\kelvin$ for clump C, which is higher than that derived from \co. At the other positions, we detected just two components, leading to very uncertain temperature estimates ($22\kel$, clump E; $45\kel$, clump B) that, however, are roughly consistent with those determined from CS (see Sect.~\ref{ssec:lvg}).
This higher temperature could be due to the presence of internal heating sources in clumps B and C.

Although the excitation temperature is formally only a measure
of the relative population of the energy levels in a transition,
and therefore differs for different transitions, 
under the assumption of LTE, it provides a fair estimate of the kinetic temperature.

\subsection{Column densities}\label{ssec:cdandmasses}

Assuming LTE conditions, the total column density of the molecular gas can be obtained from observations of the $\mathrm{J}\rightarrow\mathrm{J-1}$ transition by means of the expression  \citep{Zielinsky, KramerWinnewisser91}
\begin{equation}
\begin{split}
 N =& \frac{1}{\eta_c} \frac{3 h}{8 \pi^3 \mu^2} \frac{Z}{\mathrm{J}} \expo{\frac{\emph{h} \nu_\mathrm{J-1,0}}{\emph{k} \mathit{\tex}}} \Bigl[ 1-\expo{-\frac{\emph{h} \nu_\sbj}{\emph{k} \mathit{\tex}}} \Bigr]^{-1} [J(\tex) - J(T_\mathrm{{BG}}))]^{-1} \times \\
 \times & \int \tmb(\mathrm{J,J-1}) \mathrm{d}\upsilon \equiv f(\tex) \int \tmb(\mathrm{J,J-1}) \mathrm{d}\upsilon,
\label{eq:totcdoptthin}
\end{split}
\end{equation}
where $J(T)= (h\nu_\sbj/k)(\expo{{\emph{h} \nu_\sbj}/{\emph{k} \mathit{T}}}-1)^{-1}$, $\eta_c$ is the efficiency with which the antenna couples to the source, $\mu$ is the dipole moment of the molecule, $Z$ is the partition function, $\tbg$ is the background temperature, and $\tmb$ is the main beam temperature.

\eqn{eq:totcdoptthin} implies that there is a linear relation between the column density and the integrated line intensity at fixed $\tex$.
This equation holds only in case of negligible optical depth ($\tau$). However, from the detection equation
\begin{equation}
\tp{R}{*} = \eta_c [J(\tex)-J(\tbg)] (1-\expo{-\tau}),
\label{eq:deteq}
\end{equation}
one can derive a correction factor ${\tau}/{(1-\expo{-\tau})}$, which makes it possible to calculate the column density while $\tau \lesssim 2$, with uncertainties less than $15\%$ \citep{Kramer}. In this expression, $\tau$ is the optical depth at the line centre (see Sect.~\ref{ssec:tauav} and \ref{ssec:lvg}).

In \tablename~\ref{tab:coldens}, we list the average value of the column density inside the $3\sigma$ contour of integrated intensity, derived from different molecules and transitions, for each distinct identified component. 
The derived excitation temperature is also listed, but we note that
the column densities are computed assuming $\tex=20 \usk\kelvin$ for all molecules  and a $\tau$ derived according to Sect.~\ref{ssec:tauav}.
The assumption of a constant excitation temperature does not affect the results, because its range is small ($15-25\usk \kelvin$) and the variation in column density is only of the order of $\lesssim15-20\%$. 

\setlength{\tabcolsep}{4pt}
\begin{table*} 
\centering 
\caption{$\vlsr$, position, mean LTE column density, excitation temperature (\co), FWHM of the lines, and diameter of the clumps.}
\label{tab:coldens} 
\scriptsize
\medskip 
\begin{tabular}{cccccccc} 
\toprule 
Cl. ($\vlsr$) & Offset & $ N_{\scriptscriptstyle \mathrm{C^{18}O},10} $ & $ N_{\scriptscriptstyle \mathrm{C^{18}O},21} $  & $ N_{\scriptscriptstyle \mathrm{H_2CO},21} $ & $\tex($\co$)$ & Line FWHM & FWHM size \\ 
 (km/s) & \asec & ${(\scriptscriptstyle10^{14} \times \cm^{-2}})$ & ${(\scriptscriptstyle10^{14} \times \cm^{-2}})$ & ${(\scriptscriptstyle10^{13} \times \cm^{-2}})$ & $(\kelvin)$ & $(\kms)$ & $(\pc)$ \\ 
\midrule 
A (-6.8) & (-45,180)       &$        9.71   $&$        8.88   $&$        1.09   $&$        20   $&$        1.3  $&$        0.35  $ \\       
B (-5.6) & (-75,155)       &$        18.50  $&$        18.36  $&$        2.09   $&$        20   $&$        1.8  $&$        0.55  $ \\       
C (-5.5) & (0,40)          &$        8.14   $&$        10.10  $&$        3.79   $&$        21   $&$        2.0  $&$        0.35  $ \\      
D (-4.4) & (-70,120)       &$        23.60  $&$        20.47  $&$        2.35   $&$        20   $&$        1.8  $&$        0.56  $ \\       
E (-2.3) & (80,60)         &$        12.00  $&$        12.08  $&$        6.21   $&$        15   $&$        1.8  $&$        0.46  $ \\      
F (-2.1) & (-50,100)       &$        7.43   $&$        5.54   $&$        ...    $&$        17   $&$        1.2  $&$        0.40  $ \\       
G (-1.2) & (50,75)         &$        10.50  $&$        10.87  $&$        3.32   $&$        13   $&$        1.3  $&$        0.27  $ \\       
H (-1.2) & (-50,75)        &$        7.47   $&$        6.34   $&$        1.15   $&$        15   $&$        1.8  $&$        0.31  $ \\       
I (0.4)  & (0,95)          &$        2.81   $&$        3.12   $&$        1.14   $&$        15   $&$        1.4  $&$        0.30  $ \\       
L (-5.6) & (-75,95)        &$        5.98   $&$        3.07   $&$        ...    $&$        21   $&$        1.2  $&$        0.34  $ \\       
M (-0.3) & (-75,125)       &$        2.86   $&$        ...    $&$        ...    $&$        ...  $&$        ...  $&$        ...   $ \\       
N (-0.3) & (-50,50)        &$        2.94   $&$        ...    $&$        1.11   $&$        ...  $&$        ...  $&$        ...   $ \\       
O (-3.4) & (50,70)         &$        ...    $&$        8.07   $&$        6.56   $&$        17   $&$        1.1  $&$        0.37  $ \\       
P (-5.9) & (25,25)$^{(1)}$ &$        ...    $&$        ...    $&$        1.20   $&$        ...  $&$        ...  $&$        ...   $ \\       
\bottomrule 
\end{tabular} 
\flushleft
\textsc{Note ---} The column density is determined by averaging the emission inside the $3\sigma$ contour in $\int \tmb\mathrm{d}\upsilon$ of each molecule in each clump, assuming $\tex=20\kel$. In the first column, the mean $\vlsr$ of the clump is also indicated. The notation $ N_{\scriptscriptstyle \mathrm{A},ij} $ means that the total column density of the molecule A is derived from the transition $ij$. $^{(1)}$ The position is derived from CS(5-4).
\end{table*} 
\setlength{\tabcolsep}{5pt}

\subsection{Opacities and visual extinctions} \label{ssec:tauav}

We can derive the opacities of the transitions of \co, from the detection equation (\ref{eq:deteq}). The \co~lines are usually optically thin ($\tau\sim\pot{-2}$), although they reach $\tau \sim 0.1$ and $\tau \sim 0.2$, respectively for the transition (1-0) and (2-1) in clump E, in the elephant trunk and in some of the clumps aligned with the bright emission to the west of the trunk, i.e. the ionization front. Assuming the standard value of $\sim8$ for the abundance ratio ${X(\mathrm{^{13}CO})}/{X(\mathrm{C^{18}O})}$, one derives values of $\tau$ for the $\mathrm{^{13}CO}$ of the order of $0.8$ and $1.6$, respectively, for the transition (1-0) and (2-1) at the same positions. This result confirms that the emission of $\mathrm{^{13}CO}$ is marginally thick, as argued by \citet{Massietal97}. The higher optical depth indicates that the molecular material has accumulated in these regions. Furthermore, non-negligible opacities could explain the slightly lower $\tex$ south of the clump C, with respect to those observed for the ionization front. 

An estimate of the visual extinction can be obtained from the H$_2$ column densities from the expression \citep{Bohlinetal78}
\begin{equation}
 A_V = 5.34 \times \pot{-22} \mathrm{N_{H_2}} \usk \mg.
 \label{eq:visext870mu}
\end{equation} 
Typical values are $5-10\usk\mg$ and $15-20\usk\mg$ for \co\ and H$_2$CO, respectively. 
The maximum values of $A_V$ that we found, estimated from H$_2$CO, were those of clumps C and E ($\sim 50\usk\mg$). 

\subsection{Dust} \label{ssec:dust}

From the $870 \usk \micro \metre$ image, kindly provided by the ATLASGAL project\footnote{\url{http://www.mpifr-bonn.mpg.de/div/atlasgal/}, \citealt{Schulleretal09}}, we derived dust  masses and densities, following \citet{Deharveng09}
\begin{equation}
 \mathrm{M_d} = \frac{\mathrm{S_{870\micro\metre}} D^2}{\kappa_{870\micro\metre} \mathrm{B_{870\micro\metre}}(T_\mathrm{d})},
 \label{eq:dustmass}
\end{equation}
where $\mathrm{S_{870\micro\metre}}$ is the total flux density at $870\usk\micro\metre$, $D$ is the distance, $\mathrm{B_{870\micro\metre}}(T_\mathrm{d})$ is the emission of a blackbody with temperature equal to $T_\mathrm{d}$ at $870\usk\micro\metre$, and $\kappa_{\nu}\equiv\kappa_0(\nu/\nu_0)^\beta$ is the dust opacity per unit mass at the indicated frequency.
\citet{OssenkopfHenning94} recommend $\kappa_0=0.8\usk\cm\usk\gram^{-1}$ at $230.6\usk\giga\hertz$, from which we derived $\kappa_{870\micro\metre} \sim 1.8 \usk \cm^2 \usk \gram^{-1}$, assuming $\beta=2$ \citep{Hildebrand83}. We assumed that the dust temperature is $30\usk\kelvin$ throughout the region. However, the resulting masses and column densities do not depend strongly on temperature, for $T_\mathrm{d} = 20 \usk \kelvin$ and $T_\mathrm{d} = 50 \usk \kelvin$ the difference is a factor of three. The dust mass resulting from the total $870\mum$ flux is $\sim21\usk\msun$. 
$\kappa_{870\micro\metre}$ has an uncertainty of a factor of two, which implies an uncertainty in the dust mass of the same factor. Furthermore, the uncertainty due to the dust temperature is of the same order of magnitude.

The surface brightness $\mathrm{F_\nu}$ at $870\usk\micro\metre$ also allows one to derive the gas column density. Following \citet{Deharveng09}
\begin{equation}
 N_\mathrm{H_2} = \frac{\gamma \mathrm{F_{870\micro\metre}}}{2.3 m_\mathrm{H} \kappa_{870\micro\metre}\mathrm{B_{870\micro\metre}}(T_\mathrm{d}) \Omega_{beam}},
 \label{eq:cold870mu}
\end{equation}
$\mathrm{F_{870\micro\metre}}$ is the peak value of the surface brightness, $\Omega_{beam}$ is the solid angle covered by the beam, $m_\mathrm{H}$ is the mass of a hydrogen atom, and we assumed a gas-to-dust ratio $\gamma=100$.
The contribution of the free-free emission, extrapolated from the radio data of \citet{Fellietal90} assuming optically thin emission, is less than a few percent. The rms noise of the map ($\sim 100\usk\milli\jy\per\mathrm{beam}$) corresponds to a column density of $3.1\times\pot{21}\usk\cm^{-2}$ for $T=20\usk\kelvin$, $1.9\times\pot{21}\usk\cm^{-2}$ for $T=30\usk\kelvin$, and $1.0\times\pot{21}\usk\cm^{-2}$ for $T=50\usk\kelvin$.

\begin{figure}[tb]
 \centering
 \includegraphics[angle=-90, width=\columnwidth]{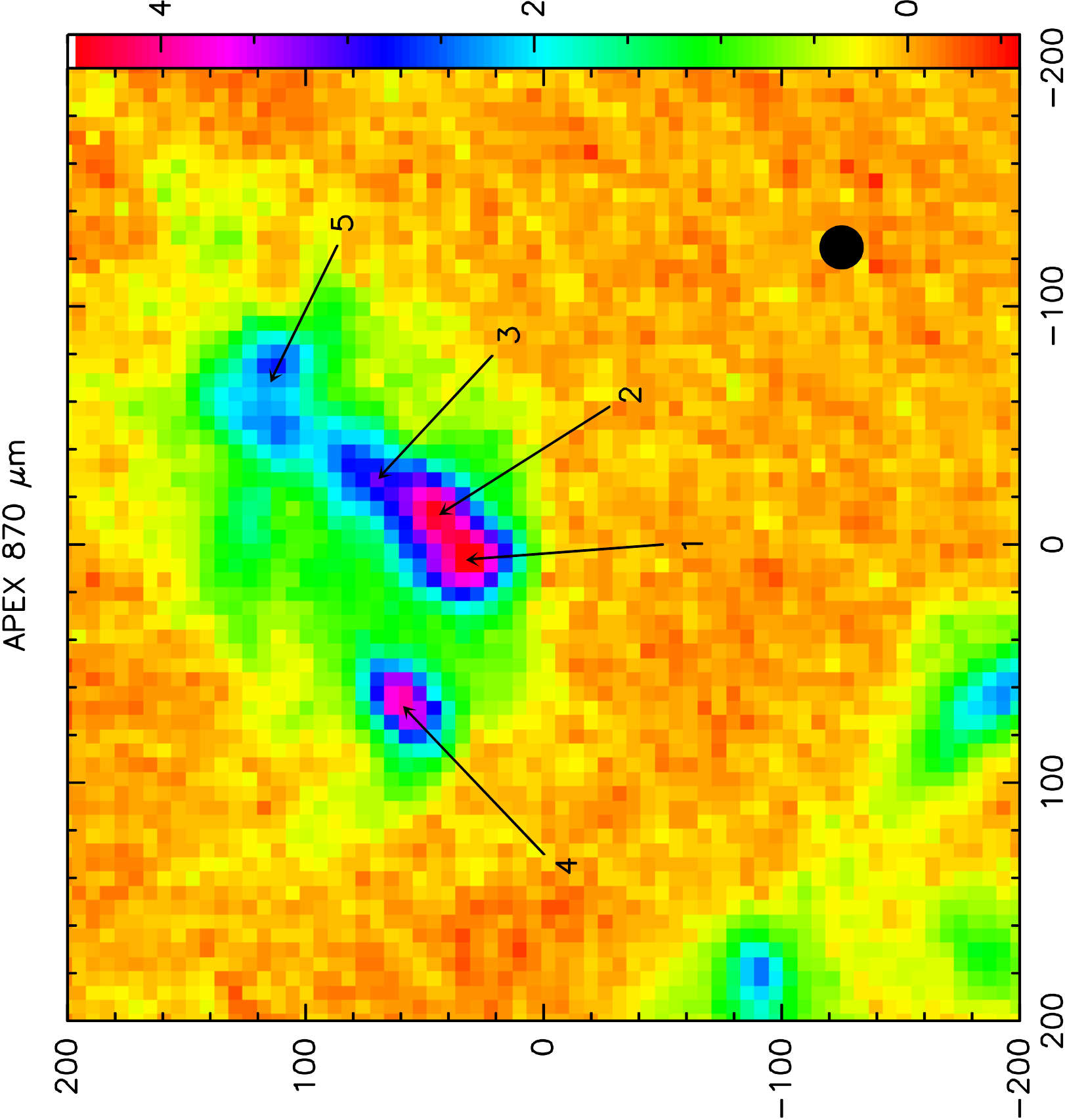}
 \caption{APEX image at $870\usk\micro\metre$. The scale is expressed in $\jy\per\mathrm{beam}$. The APEX beam is shown as a filled circle. The locations of the main components identified are shown in the map. Component 5 was not fitted with a Gaussian, but was considered as a region of diffuse emission.}
 \label{fig:contcomp}
\end{figure}

We decomposed the emission into four bi-dimensional Gaussian components with MOPSIC, plus three regions of diffuse emission: Fig.~\ref{fig:contcomp} shows the condensations identified.

The morphology of the emission at this wavelength is very similar to that of the integrated molecular emission and to that of the regions of obscuration visible at near-IR wavelengths. 
The maxima in the molecular-line emission do not always coincide with those of the dust. In particular, clump E shows a large displacement.
We suggest that this could be an effect of a non-negligible $\tau$.

\begin{figure*}[tb]
 \centering
 \includegraphics[angle=-90, width=0.75\textwidth]{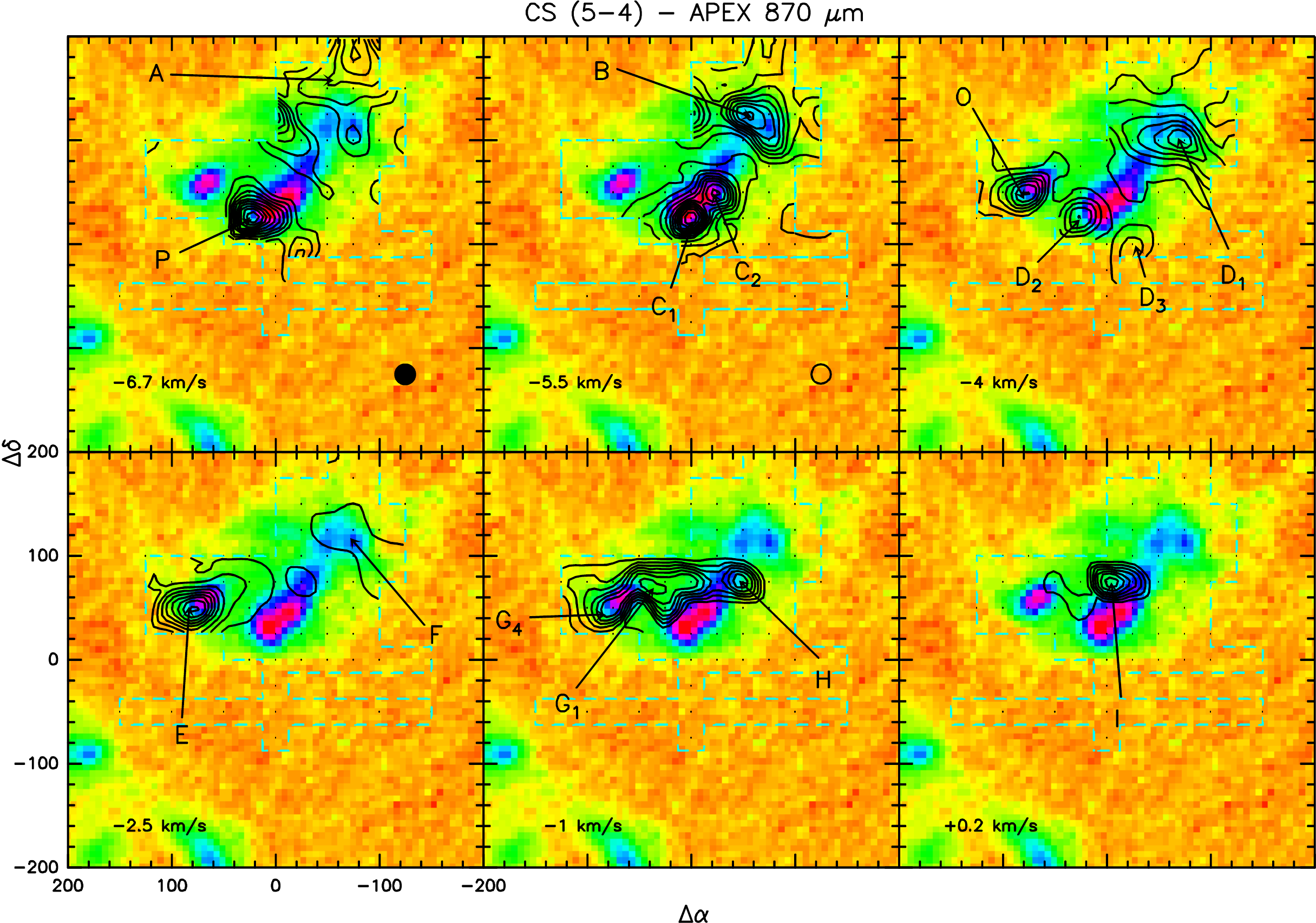}
 \caption{CS(5-4) emission (black and gray contours) superimposed on the $870\usk \mu \metre$ emission. The beam size of CS(5-4) is indicated by the filled circle, while the open circle indicates the APEX beam size. Each plot corresponds to a different $\vlsr$. The $\vlsr$ of the clumps is indicated in the respective panel. The contours (in units of $\kelvin\usk\kilo\metre\per\second$) for each clump are (lowest (step) highest): ($-6.7\usk\kilo\metre\per\second$) 0.19 (0.3) 2.59; ($-5.5\usk\kilo\metre\per\second$, C) 0.20 (1.0) 10.20; ($-5.5\usk\kilo\metre\per\second$, B) 0.19 (0.5) 3.69;  ($-4.0\usk\kilo\metre\per\second$) 0.19 (0.7) 4.39; ($-2.5\usk\kilo\metre\per\second$) 0.18 (1.0) 8.18; ($-1.0\usk\kilo\metre\per\second$) 0.17 (0.2) 1.37; and ($+0.2\usk\kilo\metre\per\second$) 0.14 (0.1) 0.74. The lowest contour corresponds to the $3\sigma$ level in $\int \tmb\mathrm{d}\upsilon$.}
 \label{fig:cs54870mu}
\end{figure*}

To associate the dust components with molecular cores, we superimposed the map of the $870 \usk \micro \metre$ emission on that of the molecular emission, as shown in Fig.~\ref{fig:cs54870mu}. 
The masses of the components identified in the APEX image and their associated molecular clumps are listed in \tablename~\ref{tab:870mumasses}. The total mass indicated in this table is computed by integrating the total $870\micro\metre$ flux. The H$_2$ column densities derived from the dust emission are on average in the range $3-7\times\pot{22}\usk\cm^{-2}$, while the maximum values are $\sim\pot{23}\usk\cm^{-2}$ (components 1, 2, and 4). 

The visual extinctions corresponding to the maximum column densities are typically $100\usk\mg$. 
The average visual extinction lies between $15$ and $40\usk\mg$, depending on the component. 
The corresponding mean and maximum volume densities, calculated assuming spherical symmetry for the clumps, are $\sim \pot{4}\usk\cm^{-3}$ and $\sim \pot{5}\usk\cm^{-3}$. These values are of the same order as those found from H$_2$CO, under the assumption of LTE. The low-density layers of the clumps are not visible in the $870\usk\micro\metre$ map owing to its high rms noise.

\begin{figure*}[tbp]
 \centering
 \includegraphics[angle=-90, width=0.9\columnwidth]{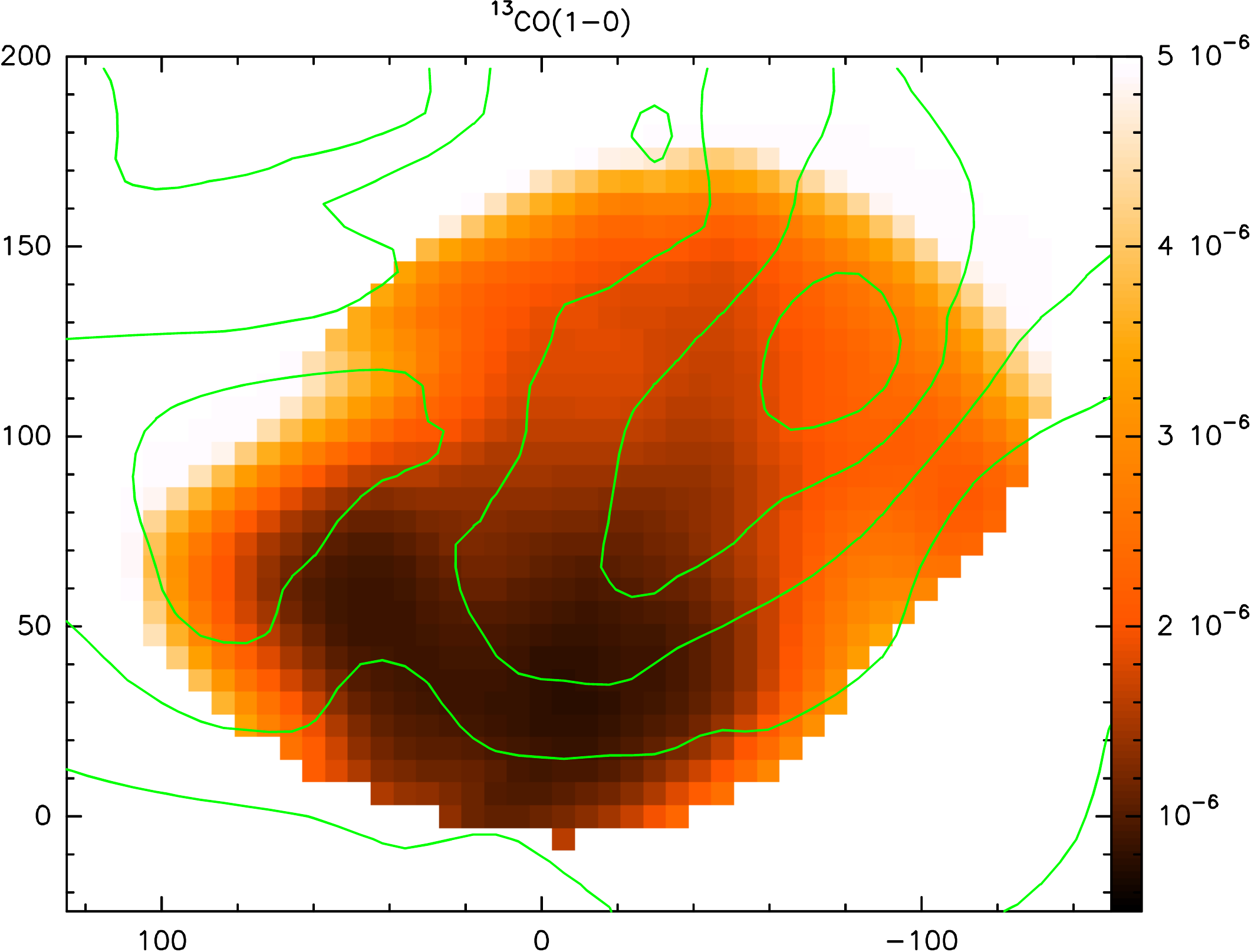}\hspace*{0.7cm}
 \includegraphics[angle=-90, width=0.9\columnwidth]{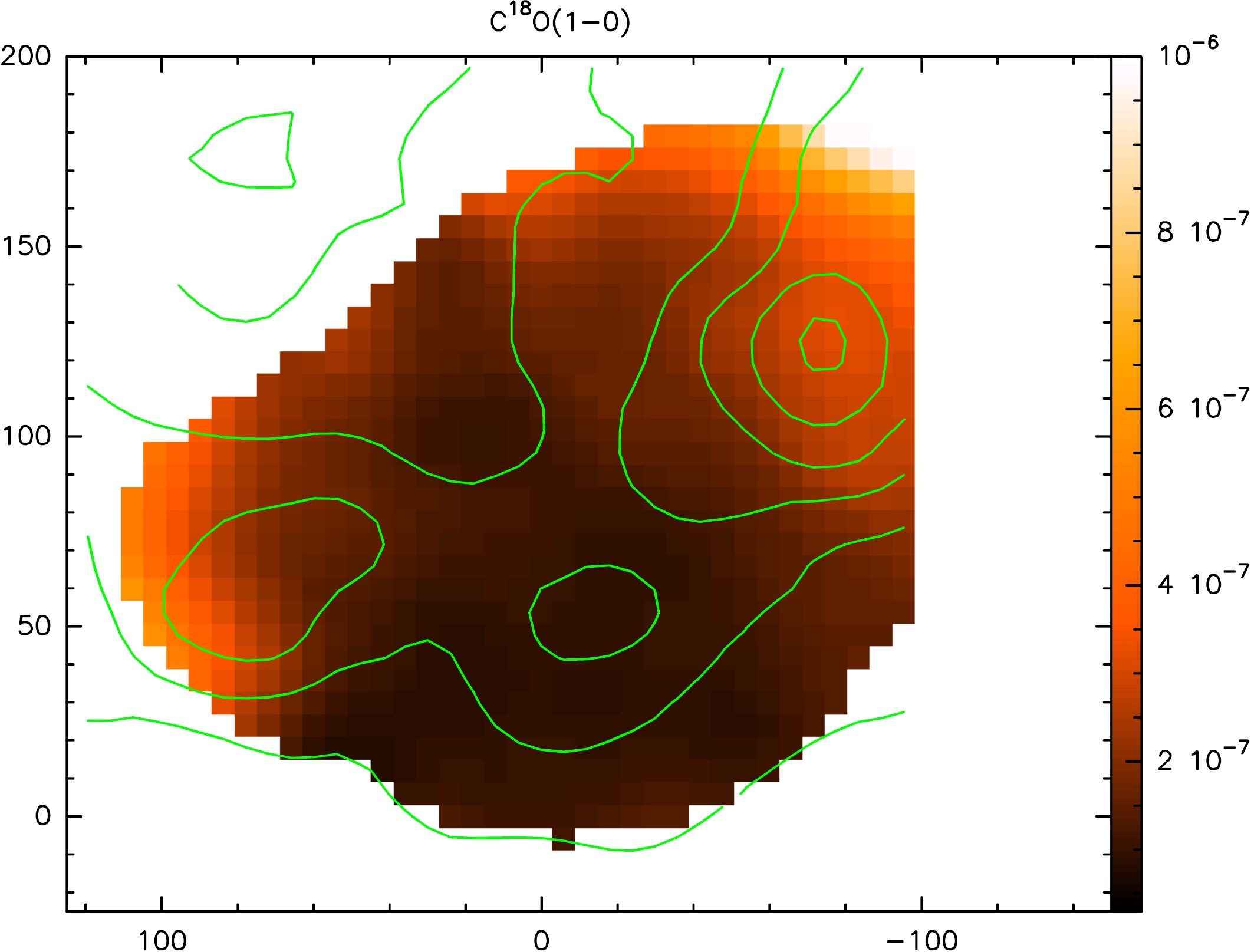}\vspace*{0.3cm}
 \includegraphics[angle=-90, width=0.9\columnwidth]{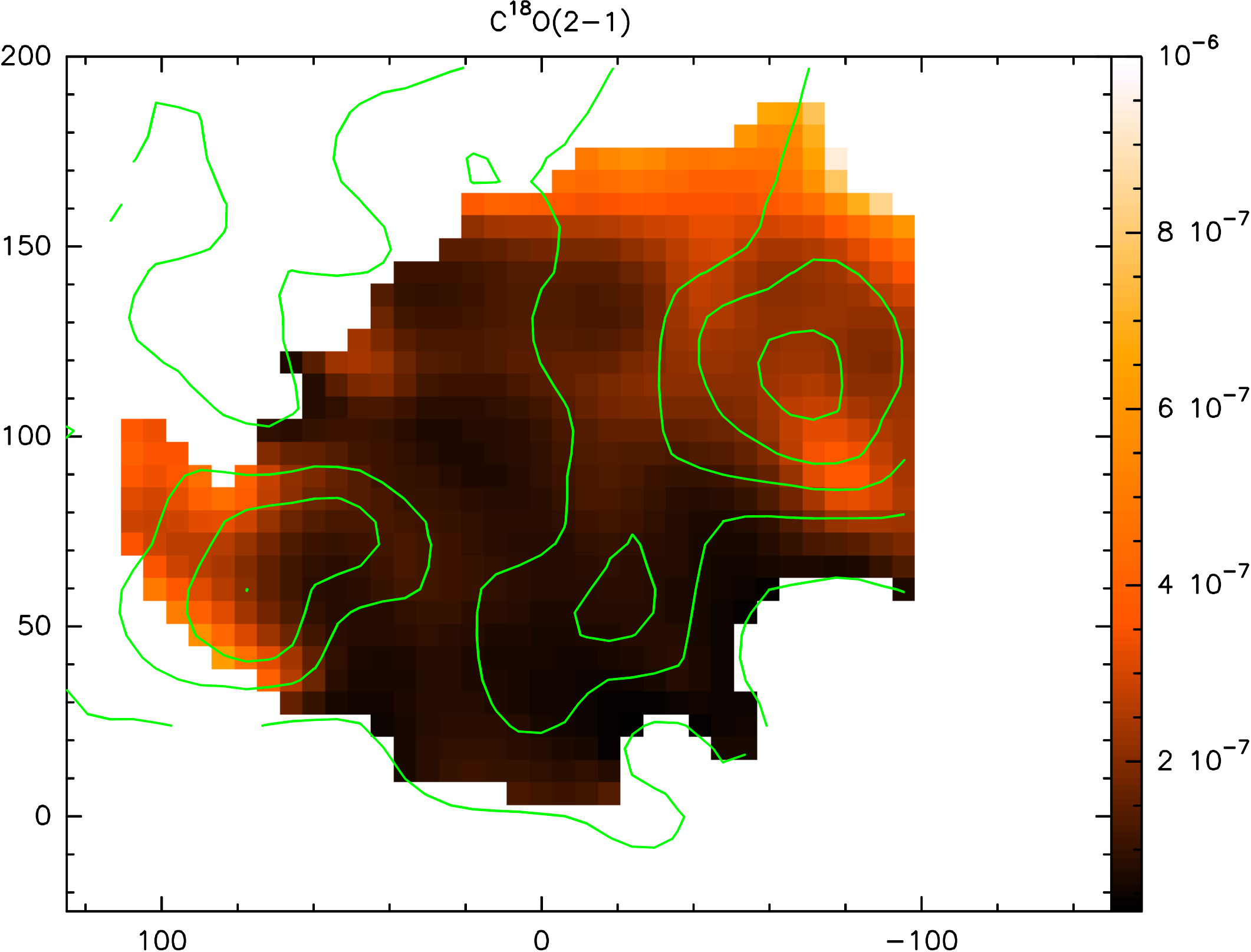}\hspace*{0.7cm}
 \includegraphics[angle=-90, width=0.9\columnwidth]{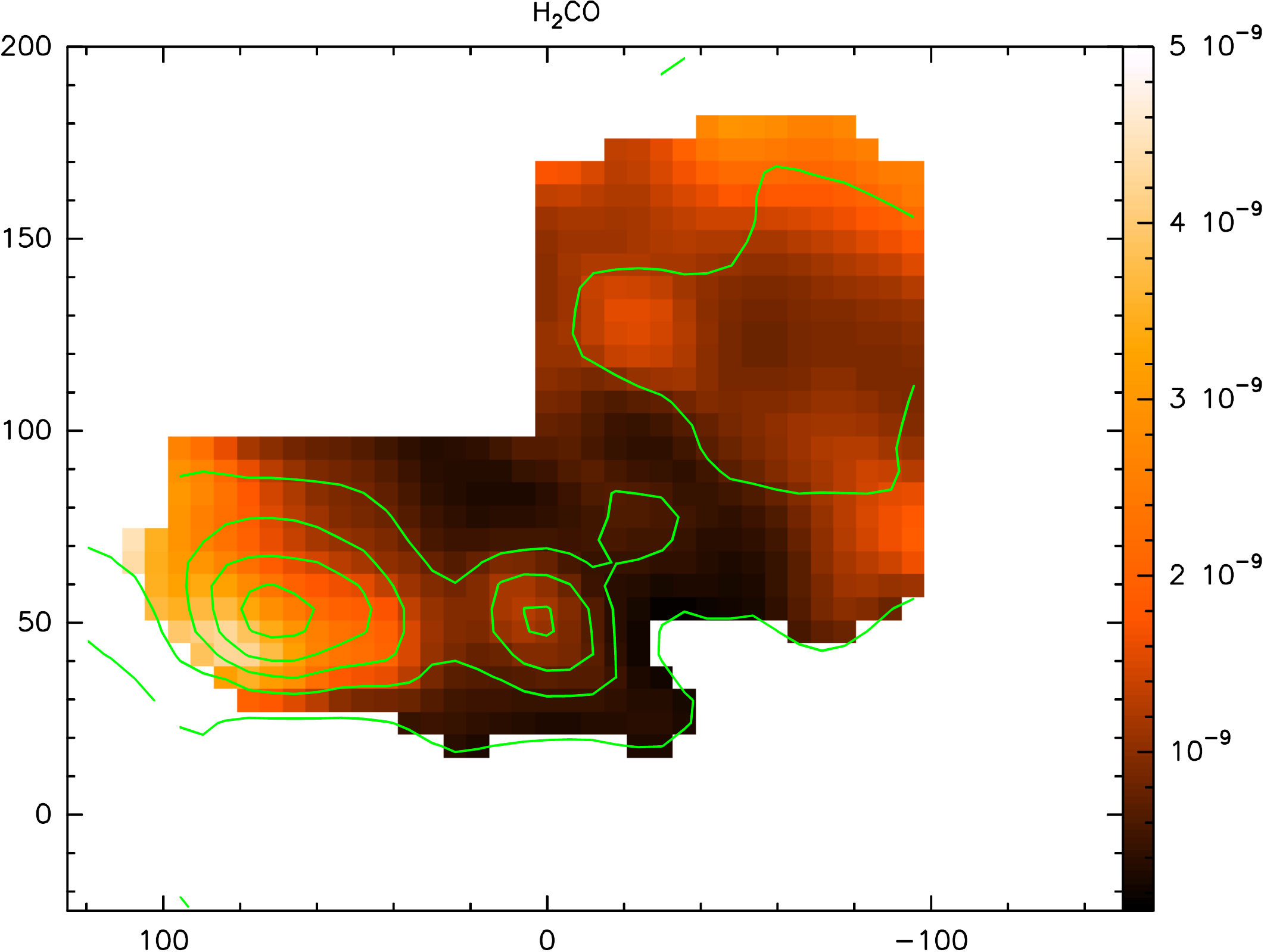}
 \caption{Maps of the abundance relative to H$_2$ of $^{13}$CO, \co, and H$_2$CO, derived from $^{13}$CO(1-0) (top left), \co(1-0) (top right), \co(2-1) (bottom left), and \formt\ (bottom right). The green contours shows the column density distribution of the molecules.}
 \label{fig:ab-map}
\end{figure*}

\subsection{Molecular abundances} \label{ssec:abund}

In Fig.~\ref{fig:ab-map}, we show the abundances of $^{13}$CO (not in the present dataset, taken from \citealt{Massietal97}), \co\ and H$_2$CO as a function of position. To derive these maps, we smoothed the APEX image to match the resolution of the molecular transition considered and we divided pixel per pixel the molecular column density map by the H$_2$ column density map, obtained from the APEX image, using Eq.~\ref{eq:cold870mu}. We set a threshold of $3\sigma$ in both molecular and H$_2$ column densities. We can clearly see that the abundances vary in the region, in a similar way for the three molecules. 
The abundances are lower near the IF and the elephant trunk, while they appear to increase further away from IF, away from Pis-24. 
The \co\ abundance varies from $\sim9.0\times\pot{-8}$ in the region of the IF, to $\sim1.9\times\pot{-7}$ at the location of E, to $\sim2.3\times\pot{-7}$ more to the north, at roughly the location of D. In the same three regions, the abundances of $^{13}$CO and H$_2$CO are $\sim1.0\times\pot{-6}$, $\sim1.5\times\pot{-6}$, and $\sim2.2\times\pot{-6}$; and $\sim6.0\times\pot{-10}$, $\sim1.5\times\pot{-9}$, and $\sim1.1\times\pot{-9}$, respectively. We expect that this variation is at least partially due to an increase in $\tk$ at the location of the ionization front.
A different $\tk$ can modify the abundance through the H$_2$ column density, which was derived from the $870\usk\mum$ emission. 
\begin{table} 
\centering 
\caption{Gas masses derived from dust emission.}
\label{tab:870mumasses} 
\scriptsize
\medskip 
\begin{tabular}{ccc} 
\toprule 
Component     & Mass    & Associated molecular clumps \\
              &$(\msun)$&                             \\
\midrule
1             & $370$   & C1                          \\
2             & $180$   & C2                          \\
3             & $210$   & H, F                        \\
4             & $340$   & E                           \\
5$^{(1)}$     & $360$   & D, A, B                     \\
total$^{(2)}$ & $2100$  &                             \\
\bottomrule 
\end{tabular} 
\\
\medskip
\textsc{Note ---} The gas-to-dust ratio was assumed to be $\gamma=100$. The total mass is computed by integrating the total $870\micro\metre$ flux. $^{(1)}$ Not a Gaussian fit, but a region that comprises all the emission at the corresponding location. $^{(2)}$ computed integrating the whole $870\mu$ flux. \\
\end{table}

\subsection{LTE masses and volume densities}

The masses of the clumps are listed in Table~\ref{tab:totmasses}. They were derived averaging the emission inside the $3\sigma$ contour to determine the average column density and then integrating over the beam-corrected area of the emission. We also took into account the optical depth $\tau$ (determined from the detection equation) and a correction for helium, which contributes with a factor $1.36$. 
The total mass of the complex is $\sim 2000 \usk \solm$, in agreement with that found from the $870\mum$ continuum emission. This mass can be compared to the total mass of ionized hydrogen ($290\usk\msun$, \citealt{Bohigasetal04}) and to the total dust mass ($21\usk\msun$, see Sect.~\ref{ssec:dust}). The masses of single clumps range from a few tens to several hundreds of $\solm$.

\begin{table} 
\centering 
\caption{Single-clump and total gas masses.} 
\label{tab:totmasses} 
\scriptsize
\medskip 
\begin{tabular}{cccc} 
\toprule 
Cl. & ${\scriptscriptstyle \mathrm{C^{18}O (1-0)}}$ & ${\scriptscriptstyle \mathrm{C^{18}O (2-1)}}$ & ${\scriptscriptstyle \mathrm{H_2CO (2_{1,2}-1_{1,1})}}$  \\
 & $(\msun)$ & $(\msun)$ & $(\msun)$ \\ 
\midrule 
A     &$        130  $&$        160  $&$        170  $ \\
B     &$        320  $&$        310  $&$        330  $ \\
C     &$        270  $&$        270  $&$        350  $ \\
D     &$        570  $&$        400  $&$        300  $ \\
E     &$        210  $&$        240  $&$        300  $ \\
F     &$         70  $&$         50  $&$        140  $ \\
G     &$        130  $&$         90  $&$         60  $ \\
H     &$        180  $&$        170  $&$        250  $ \\   
I     &$         20  $&$         30  $&$         60  $ \\
L     &$         20  $&$         40  $&$        ...  $ \\
M     &$       <10   $&$        ...  $&$        ...  $ \\
N     &$         20  $&$        ...  $&$         20  $ \\
O     &$        ...  $&$         50  $&$         90  $ \\
P     &$        ...  $&$        ...  $&$         90  $ \\
\midrule
Total & $1950$        & $1810$        & $2090$         \\
\bottomrule 
\end{tabular} 
\\
\medskip
  \textsc{Note ---} The mass is computed integrating the column density within the $3\sigma$ contour in $\int \tmb\mathrm{d}\upsilon$ for each clump. The transition from which we derived the mass is indicated in the top row. If the clump is divided into sub-clumps (indicated by numbers; cf. Fig.~\ref{fig:cs-c18o}) the mass reported is the sum of all the sub-components.
\end{table}

The volume density $n$ is determined assuming that the extent of the clump along the line-of-sight is equal to its size in the plane of the sky. The ratio of volume densities determined from low- and high-density tracers may be up to a factor of $\sim 100$, e.g., the mean volume densities derived from \co\ are typically $\sim\pot{3} \usk \centi \metre ^{-3}$, while those of H$_2$CO are $\approx \pot{4}-\pot{5} \usk \centi \metre ^{-3}$. The volume densities derived in this way are always of the same order of magnitude of the critical density of the molecular transition used to determine it.

\subsection{Non-LTE analysis} \label{ssec:lvg}

The three transitions of CS, plus C$^{34}$S(2-1), allowed a non-LTE analysis at offsets where data points were available for all transitions. For this purpose, we used the statistical equilibrium, radiative transfer code RADEX\footnote{\url{http://www.strw.leidenuniv.nl/$\sim$moldata/radex.html}; \citealt{VanderTaketal07}}, for the approximation of a uniform sphere. The model requires as input the kinetic temperature, the H$_2$ number density, the molecular column density,  and the FWHM of the line. We assumed the Solar value of $22.5$ for the $^{32}$S/$^{34}$S isotopic ratio. The model returns the brightness temperature of the lines, the opacity, and the excitation temperature. The FWHM of the CS line was determined by averaging those of the various observed lines, at one position. The brightness temperature of the line was estimated using 

\begin{equation}
 \tb = \frac{\vartheta_\mathrm{B}^2+\vartheta_\mathrm{S}^2}{\vartheta_\mathrm{S}^2} \tmb,
\end{equation}
where $\vartheta_\mathrm{B}$ is the beam size and $\vartheta_\mathrm{S}$ is the source size. We assumed that all the transitions come from the same region, estimating the source size from the mean FWHM dimension of clumps in CS(5-4), that has the highest angular resolution. We varied the kinetic temperature between $10\usk\kelvin$ and $200\usk\kelvin$, and the molecular hydrogen number density between $\pot{3}\usk\cm^{-3}$ and $\pot{7}\usk\cm^{-3}$, both in $50$ equally spaced logarithmic steps. The column density was varied between $5.2\times\pot{12}\usk\cm^{-2}$ and $5.5\times\pot{15}\usk\cm^{-2}$ for CS and between $2.3\times\pot{11}$ and $2.4\times\pot{14}\usk\cm^{-2}$ for C$^{34}$S, in $44$ equal logarithmic steps. 

To analyze the model results, we used a Bayesian approach. The Bayes theorem states that

\begin{equation}
P(\tk,N,n|D,m) = \frac{1}{\varphi} P(D|m,\tk,N,n)\, P(\tk,N,n),
\label{eq:bayestheo}
\end{equation}
where $P(\tk,N,n|D,m)$ is the probability of the parameters $\tk,N$, and $n$, given the data and the model, called \textit{posterior}, $P(D|m,\tk,N,n)$ is the probability of the data given the model and its parameters, or \textit{likelihood}, and $P(\tk,N,n)$ is the probability of the model parameters, which is called \textit{prior}. The parameter $\varphi$ is a normalization constant given by the sum of the individual probability of each model, in order to have the posterior normalized to 1. We used a constant prior for the model parameters, thus giving equal weights to every value of the model parameters. The probability of measuring a certain value for the intensity of a line is assumed to be represented by a Gaussian curve centred on the value obtained from RADEX for specific physical conditions of the gas, and with a $\sigma$ given by the uncertainty in the measured value. Therefore, we multiplied four Gaussian curves, one for each transition, and the probability density function (PDF) for $\tau_\mathrm{C^{34}S}$, computed with JAGS\footnote{\url{http://mcmc-jags.sourceforge.net/}} assuming that CS(2-1) is optically thick, to obtain $P(D|m,\tk,N,n)$. The assumption of optical thickness for CS(2-1) is only used to derive the PDF for $\tau_\mathrm{C^{34}S}$, but is not used further in the analysis of the model results. 
Explicitly, the expression for $P(D|m,\tk,N,n)$ is

\begin{equation}
 P(D|m,\tk,N,n) = \frac{1}{\varphi} \Bigl[ \prod^4_{i=1} (\expo{-(\emph{l}_i-\mu_i)^2/(2\sigma_{l,i}^2)}) \Bigr] P(\tau_\mathrm{C^{34}S}),
\label{eq:likelihood}
\end{equation}
where the index $i$ runs over the three CS lines and C$^{34}$S(2-1), $l_i$ are the observed line intensities, $\mu_i$ are the modeled intensities, $\sigma_{l,i}$ takes into account the rms of the spectrum and a $15\%$ calibration uncertainty, and $\varphi$ is a normalization constant.

To determine the physical conditions of the gas, i.e. the single parameters of the model, we had to integrate over the other parameters (\textit{marginalize})

\begin{equation}
P(a|D,m)=\int P(a,b,c|D,m) \,db\, dc.
\label{eq:marginalization}
\end{equation}

From the PDF of the parameters, we derived the expectation values and the $1\sigma$ range.
The results for CS are summarized in Table~\ref{tab:radex-cs}. The columns shows the offset of the spectrum used, the clump name, the kinetic temperature and its $1\usk\sigma$ range, the number density of molecular hydrogen and its $1\usk\sigma$ range, the column density and its $1\usk\sigma$ range, the molecular abundance, the optical depth of CS(2-1), (3-2), (5-4) and C$^{34}$S(2-1), and the excitation temperature of the same transitions, respectively.

From these analyses, we found that the volume density of molecular hydrogen ranges between $\mathrm{several} \times \pot{4}$ and $\mathrm{few} \times \pot{6}\cm^{-3}$, while $\tk$ lies between $\sim 11 \usk \kelvin$ and $\sim 45 \usk \kelvin$, for different clumps. The temperature derived from \metac\ for clump C is higher than the kinetic temperature obtained from this analysis for (0\asec,50\asec), while it is consistent with that at (0\asec,0\asec), most probably because the \metac\ beam takes in the emission from the region along the IF. However, CS(5-4) has its maxima at (0\asec,25\asec) and at (25\asec,50\asec) where we do not have data for CS(2-1) and (3-2), while at (0\asec,50\asec) the emission of CS(5-4) is weak.

There is a clear trend showing an increase in the densities towards the south, in the direction of Pis-24 and in the clumps aligned with the IF (Fig.~\ref{fig:kbandif}) west of the elephant trunk. This region shows typical densities of $\mathrm{few} \times \pot{5}\usk\cm^{-3}$ and $\tk\sim30-40\usk\kelvin$. The temperature for the point at (0\asec,0\asec) appears to be higher than the surrounding points, with $\tk\sim 40 \usk\kelvin$, while (0\asec,50\asec) has $\tk\sim 18 \usk\kelvin$.
The column density ranges between $\sim 2\times\pot{13}\usk\cm^{-2}$ and $\sim 2\times\pot{14}\usk\cm^{-2}$, implying CS abundances of between $7.0\times\pot{-10}$ and $4.0\times\pot{-8}$, derived from the ratio with the H$_2$ column densities determined from the APEX data.
The opacity of CS(2-1), CS(3-2) and CS(5-4) lies between $0.5-4$, $0.8-6$, and $0.1-2.7$, respectively. 

The derived $\tex$-values are around $7-16\usk\kelvin$ for CS(2-1), $6-11\kel$ for CS(3-2), $5-9\kel$ for CS(5-4) and $6-13\kel$ for C$^{34}$S(2-1). The ratios of $\tex$ for different transitions are always within a factor of two of each other. 

\setlength{\tabcolsep}{4pt}
\begin{table*}[tbp] 
\centering 
\caption{Summary of RADEX results from CS data, for selected offsets.}
\label{tab:radex-cs} 
\scriptsize
\medskip 
\begin{tabular}{ccccccccccccccccc} 
\toprule 
Offset     & Cl. & $\tk$     & $1\sigma$ & $n_\mathrm{H_2}$    & $1\sigma$               & $N_\mathrm{CS}$         & $1\sigma$                & [CS/H$_2$]     & $\tau_{21}$ & $\tau_{32}$ & $\tau_{54}$ & $\tau_\mathrm{C^{34}S,21}$ & $\tex$ $_{,21}$ & $\tex$ $_{,32}$ & $\tex$ $_{,54}$ & $\tex$ $_{C^{34}S,21}$ \\
(\asec)      &     & $(\kelvin)$ & $(\kelvin)$ & $(\pot{4}\times\cm^{-3})$ & $(\pot{4}\times\cm^{-3})$ & $(\pot{13}\times\cm^{-2})$ & $(\pot{13}\times\cm^{-2})$ & $(\pot{-9})$ &             &             &             &                 & $(\kelvin)$   & $(\kelvin)$ & $(\kelvin)$ & $(\kelvin)$   \\ 
\midrule 
(-100,200) & A   & $33$      & $10-64$   & $17$                & $4.8-36$                & $4.2$                    & $3.9-4.9$                & $26$       & $0.7$       & $1.5$       & $0.5$                        & $0.05$              & $14$        & $10$        & $7$           & $12$        \\
(-50,150)  & B   & $25$      & $11-35$   & $18$                & $8.3-30$                & $7.4$                    & $6.1-7.7$                & $5.2$      & $1.4$       & $2.5$       & $1.2$                        & $0.09$              & $13$        & $10$        & $6$           & $11$        \\
(0,0)      & C   & $40$      & $10-72$   & $5.3$               & $1.9-8.3$               & $4.3$                    & $3.1-6.1$                & $9.5$      & $1.2$       & $1.4$       & $0.1$                        & $0.08$              & $8$         & $6$         & $6$           & $6$         \\
(0,50)     & C   & $18$      & $11-20$   & $25$                & $14-52$                 & $21$                     & $14-52$                  & $2.7$      & $2.4$       & $4.1$       & $2.2$                        & $0.17$              & $13$        & $11$        & $6$           & $10$        \\
(-50,100)  & D   & $26$      & $10-40$   & $22$                & $5.8-63$                & $16$                     & $15-19$                  & $3.4$      & $1.6$       & $3.1$       & $2.0$                        & $0.12$              & $16$        & $12$        & $7$           & $13$        \\
(100,50)   & E   & $11$      & $10-15$   & $260$               & $130-480$               & $18$                     & $15-19$                  & $40$       & $3.8$       & $5.1$       & $2.7$                        & $0.20$              & $11$        & $11$        & $9$           & $11$        \\
(-100,100) & F   & $29$      & $10-40$   & $14$                & $5.8-25$                & $2.1$                    & $1.6-2.5$                & $0.7$      & $0.5$       & $0.8$       & $0.1$                        & $0.03$             & $11$        & $7$         & $6$           & $10$        \\
\bottomrule 
\end{tabular} 
\\
\medskip
\end{table*}
\setlength{\tabcolsep}{5pt}

\setlength{\tabcolsep}{4pt}
\begin{table*}[tbp] 
\centering 
\caption{Summary of RADEX results from CN data, for selected offsets.}
\label{tab:cn-radex} 
\scriptsize
\medskip 
\begin{tabular}{ccccccccccccc} 
\toprule 
Offset     & Cl. & $\tk$     & $1\sigma$ & $n_\mathrm{H_2}$    & $1\sigma$               & $N_\mathrm{CN}$         & $1\sigma$                & $\tau_{10}$ & $\tau_{21}$ & $\tex$ $_{,10}$ & $\tex$ $_{,21}$ \\
(\asec)      &     & $(\kelvin)$ & $(\kelvin)$ & $(\pot{5}\times\cm^{-3})$ & $(\pot{5}\times\cm^{-3})$ & $(\pot{14}\times\cm^{-2})$ & $(\pot{14}\times\cm^{-2})$ &          &              & $(\kelvin)$   & $(\kelvin)$ \\ 
\midrule 
(-50,150)  & A   & $35$      & $21-57$   & $2.8$               & $1.2-7.6$               & $2.6$                    & $2.2-3.4$                & $1.2$      & $4.0$       & $17$        & $10$                                               \\
(-75,150)  & B   & $25$      & $11-35$   & $2.3$               & $0.8-7.6$               & $0.7$                    & $0.6-0.9$                & $0.9$      & $1.6$       & $9$         & $6$                                                \\
(0,50)     & C   & $33$      & $26-40$   & $18$                & $11-28$                 & $2.3$                    & $1.9-3.0$                & $0.2$      & $1.6$       & $67$        & $18$                                              \\
(-50,100)  & D   & $45$      & $30-68$   & $3.3$               & $1.7-6.3$               & $2.2$                    & $1.6-3.0$                & $0.7$      & $3.2$       & $26$        & $11$                                               \\
(75,50)    & E   & $34$      & $28-42$   & $37$                & $23-69$                 & $1.8$                    & $1.4-2.2$                & $0.2$      & $1.1$       & $62$        & $25$                                               \\
(50,75)    & G   & $41$      & $28-58$   & $6.8$               & $3.0-16$                & $1.9$                    & $1.4-2.5$                & $0.2$      & $2.2$       & $62$        & $14$                                               \\
(-50,100)  & H   & $33$      & $21-57$   & $2.5$               & $1.0-6.3$               & $1.7$                    & $1.4-2.2$                & $1.2$      & $3.5$       & $15$        & $8$                                                \\
\bottomrule 
\end{tabular} 
\\
\medskip
\end{table*}
\setlength{\tabcolsep}{5pt}

\medskip

A similar analysis was carried out for CN. In this case, $P(D|m,\tk,N,n)$ was calculated by comparing total fluxes rather than line temperatures, due to hyperfine splitting, which is not included in the present model. Therefore we have

\begin{equation} 
 P(D|m,\tk,N,n) = \frac{1}{\varphi} \Bigl[ \prod^2_{i=1} (\expo{-({F_i}-F_{m,i})^2/(2\sigma_{F,i}^2)}) \Bigr] \expo{-(\tau_{tot,(1-0)}-\tau_{m,(1-0)})/(2\sigma_\tau^2)},
\label{eq:likelihood-cn}
\end{equation}
where $F_i$ is the measured flux, $F_ {m,i}$ is the output of the model, $\tau_{tot,(1-0)}$ is the optical depth of the (1-0) transition as measured from the hyperfine satellite ratios, and 
$\tau_{m,(1-0)}$ is that predicted by the model. The uncertainty $\sigma_{F,i}$ takes into account the rms of the integral and a $15\%$ calibration uncertainty, and $\sigma_\tau$ is the uncertainty in $\tau_{tot}$, as given by CLASS. We used a Gaussian prior on $\tk$, centred on $35\kel$, with a $\sigma_{T_K}=30\kel$, given the results of the analysis carried out for CS and the CO results in \citet{Massietal97}. The parameter $\varphi$ is the normalization constant. Also in this case, the column and the number densities are constrained quite well, while the temperature is much more uncertain. 

The results obtained from RADEX for CN are summarized in Table~\ref{tab:cn-radex}. The columns shows the offset of the spectrum used, the clump name, the kinetic temperature and its $1\usk\sigma$ range, the number density of molecular hydrogen and its $1\usk\sigma$ range, the column density and its $1\usk\sigma$ range, the molecular abundance, the optical depth of CN(1-0) and (2-1), and the excitation temperature of the same transitions, respectively. The values listed in Table~\ref{tab:cn-radex} for $\tau_{10}$ are in very good agreement with those derived from the observations.

Owing to the poor constraints on $\tk$, temperature maps are the most difficult to interpret. However, we obtain typical temperatures of between $25\kel$ and $32\kel$ for the whole region. 
The difference in $\tk$ found with CS at (0\asec,50\asec) ($18\kel \,vs.\, 33\kel$) might be understood if one considers that CN is a good tracer of PDRs \citep{Simon97}, thus the emission may come from the more external material, directly heated and shocked by the interaction with the early-type stars of Pis-24. In all cases, these temperatures are usually within the $1\sigma$ interval also found for E and at the edges of D and  B, nearly facing Pis-24.

In contrast, we derived a slightly lower temperature ($\sim25\kel$) for C2 than for C1, which is nevertheless consistent with those cited above.

The clumps in the region have a typical number density in the range $\sim1-6\times\pot{5}\usk\cm^{-3}$. The number density is highest along the IF (C1 and C2), in D and in E, with values up to a $\mathrm{few}\times \pot{6}\usk\cm^{-3}$. Clump E is particularly interesting, since it  appears to have a resolved compressed layer facing Pis-24. The H$_2$ number densities that we find are on average higher than those derived from CS, consistent with the idea that most of the emission comes from the high-density surface layers of the PDR.

\begin{figure}[tbp]
 \centering
 \includegraphics[angle=-90, width=0.9\columnwidth]{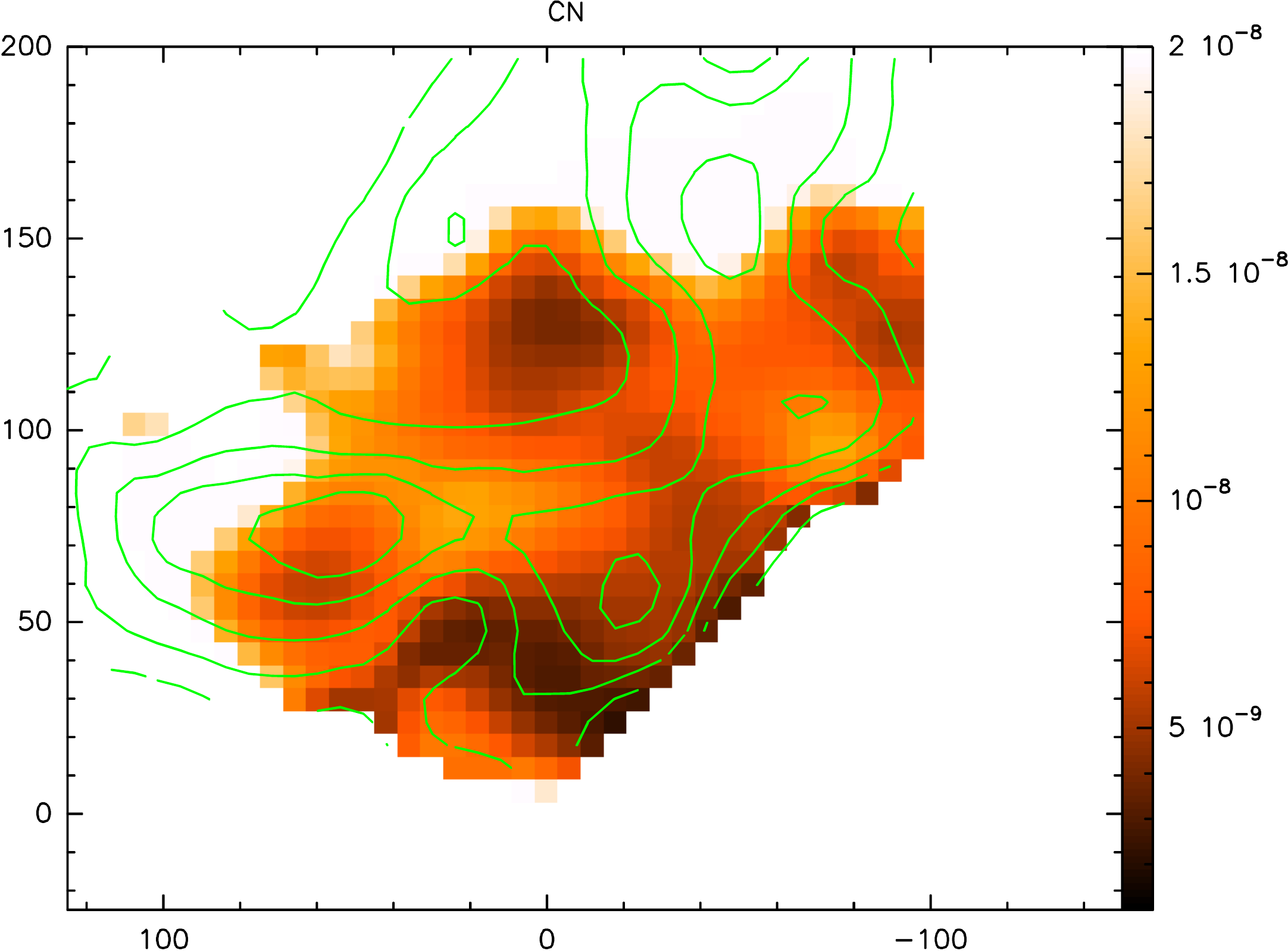}
 \caption{Map of molecular abundance relative to H$_2$, derived from CN data, through a non-LTE analysis. The green contours shows representative values of the column density, to clarify the distribution of the molecule.}
 \label{fig:cn-ab-map}
\end{figure}

The column density of single clumps lies between $\sim 5.0 \times \pot{13}\cm^{-2}$ and $\sim 3.0 \times \pot{14}\cm^{-2}$.

After determining the column density for the various components, we were able to construct an integrated column density map, hence derive an abundance map similar to that of $^{13}$CO, \co, and H$_2$CO, as given in Fig.~\ref{fig:cn-ab-map}. The emerging abundance pattern closely resembles that of the other three molecules, indicating once more that the region of the elephant trunk and the brightest part of the IF is where the influence of Pis-24 is the strongest. The abundance of CN is $\sim5.0\times\pot{-9}$ in the region of C, $\sim8.0\times\pot{-9}$ around D, and $\sim7.9\times\pot{-9}$ around E. We found that CN is enhanced around clump C and the IF, where the NIR emission is at its strongest and the ionizing radiation more intense: this is suggested by a less pronounced variation in abundance than that found from other molecules. For CN, the column density peak of the gas and that of the dust for clump E coincide, showing that its displacement for the other molecules can be caused by optical depth effects (cf. Sect.~\ref{ssec:abund}) or that \co\ and H$_2$CO are frozen onto grains at the centre of the clump (given also the very low temperature derived from CS). However, a displacement is observed for the peak in the region of clumps A and B, again toward Pis-24. This and the temperatures comparable to those of C1 might indicate that the IF/PDR extends here, even though the brightness at NIR wavelengths is much lower than nearer Pis-24. 

\subsection{Virial masses}

The total mass of a spherical system in virial equilibrium is given by \citep{MacLaren88}
\begin{equation}
 M_\mathrm{{vir}}[\msun]=k_2 R[\pc] \Delta \upsilon[\kilo\metre\per\second]^2,
 \label{eq:virmassastro}
\end{equation}
where $R$ is the radius and $\Delta \upsilon^2=8 \ln 2 \sigma^2$, assuming a Gaussian velocity profile and a density profile described by a power law $\rho(r) \propto R^{-q}$ with $q < 3$. 
The values of $k_2$ are $210$, $190$, and $126$, respectively for $q = 0,1,2$ \citep{MacLaren88}. The uncertainty caused by the unknown density profile is approximately a factor of two.
This expression neglects the influence of magnetic fields, rotation, and internal energy sources, which are usually non-negligible in molecular clouds.

To estimate $R$ for Eq.~\ref{eq:virmassastro}, we used the ``effective radius'', i.e. the radius of a circle with the same area as the clump above the FWHM level, corrected for the beam size. When the FWHM is smaller than the beam size, we assumed as an upper limit to the angular size, half of the actually observed FWHM.
As a consequence of the large beam, some individual clumps may be blended and appear as one. For example, C is resolved into two different clumps in CS(5-4), and less clearly also in \co(2-1), while it appears to be unresolved in the other transitions. 

\begin{figure}[tbp]
 \centering
 \includegraphics[angle=-90, width=\columnwidth]{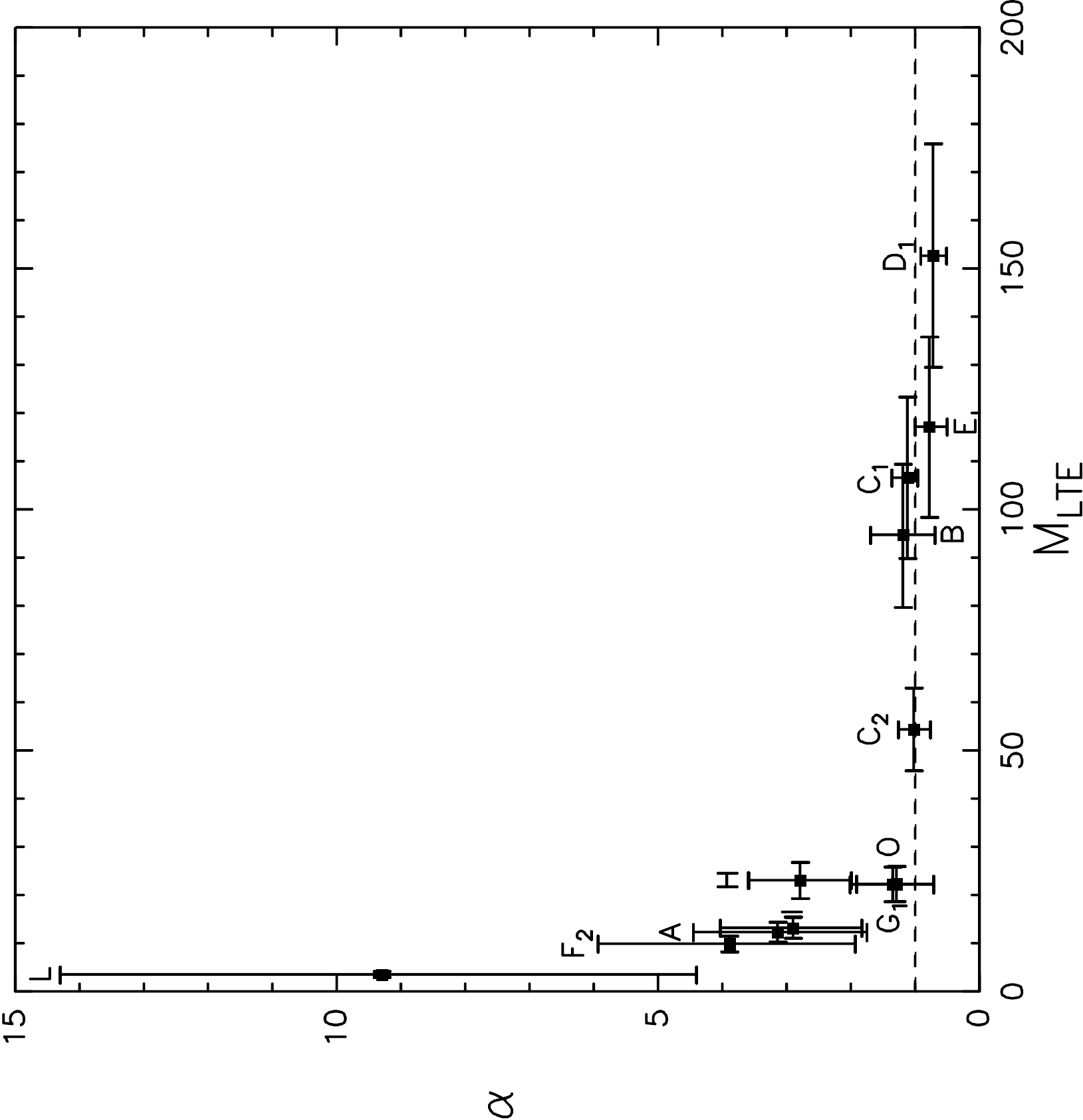}
 \caption{Virial parameter $\alpha$ as a function of $M_\mathrm{LTE}$. $M_\mathrm{vir}$ and $M_\mathrm{LTE}$ are both determined from \co. $M_\mathrm{LTE}$ is calculated within the FWHM contour in $\int \tmb \mathrm{d} \upsilon$. The dashed line indicates  $\alpha\sim1$, i.e. $M_\mathrm{vir}=M_\mathrm{LTE}$.}
 \label{fig:alphamlte}
\end{figure}

Figure~\ref{fig:alphamlte} shows the virial parameter $\alpha=M_\mathrm{vir}/M_\mathrm{LTE}$ as a function of $M_\mathrm{LTE}$, both determined from \co(2-1), where we assumed that $q=2$.  The mass $M_\mathrm{LTE}$ was derived within the FWHM contour in $\int \tmb \mathrm{d}\upsilon$. 
All clumps with masses above $50\usk\msun$, and also two with lower masses ($M\sim 20\usk\msun$ G1 and O) have $\alpha\approx 1$, thus indicating that these clumps might be gravitationally bound.
We note, however, that this is an oversimplification of the problem of stability and must be taken with caution.

\subsection{Selective photodissociation}

G353.2+0.9 is illuminated by several early-type stars. These stars deliver a huge quantity of energetic photons that dominate the chemistry and the heating of the gas in the PDR. The incident far-ultraviolet (FUV) flux is usually measured in terms of the ratio of the FUV flux in the region and the mean interstellar flux (the Habing Field, $1.6\times\pot{-3}\usk\erg\usk\cm^{-2}\usk\second^{-1}$). 

Considering the luminosities of the O3.5 If* [$\log(L/L_\odot)\sim6$], of the O3.5 III(f*) [$\log(\mathrm{L}/\mathrm{L_\odot})\sim5.9$], and of the O4 III(f+) [$\log(L/L_\odot)\sim5.8$] stars given by \citet{WeidnerVink10}, we can assume that these stars dominate the emission of energetic photons in the region. The fraction of luminosity emitted in the FUV band was estimated using a simple black-body law, between $912\usk\angstrom$ and $2067\usk\angstrom$. We obtained a total FUV luminosity of
\begin{equation}
 \mathrm{L_{FUV,tot}} = 1.2\times\pot{6}\usk \mathrm{L_\odot},
\end{equation}
and making use of
\begin{equation}
 G_0 = \frac{1}{1.6\times\pot{-3}\usk\erg\usk\cm^{-2}\usk\second^{-1}}\frac{\mathrm{L_{tot}}}{4\pi D^2},
\end{equation}
we found that $G_0\sim5.6\times \pot{4}$, for a representative projected distance of the stars from the elephant trunk of $0.66\usk\pc$. On the other hand, for the ionization front, we obtained $G_0\sim2.0\times \pot{4}$, using a projected distance of $1.12\usk\pc$ between the IF and the cluster.

The high FUV flux may influence the ratio of \co\ and $^{13}$CO, by means of selective photodissociation. This process tends to destroy the less abundant isotopologue, owing to the lower optical depth, thus less effective self-shielding capability.

We obtained the ratio of the transitions of the two isotopologues ($^{13}$CO$/$\co) for both observed transitions, combining $^{13}$CO and \co\ values measured in positions less than
half a beam apart. We corrected for the effects of opacity of the $^{13}$CO, evaluated through the detection equation (given in Eq.~\ref{eq:deteq}). The ratio of the $\tmb$ of the lines should therefore be representative of the relative abundance of the two molecules.
The $^{13}$CO$/$\co\ ratio is everywhere consistent with the standard value of the relative abundance for the two isotopologues ($\sim 8$), with three exceptions.
The measured values of the $^{13}$CO$/$\co\ ratio south of the ionization front, are a factor of two  higher than the standard value. South of the elephant trunk, we did not detect \co. However, assuming emission from \co\ at rms intensity we obtain a lower limit to the ratio of $7-10$. 
The model of \citet{Visseretal09} shows that the relative increase in the $^{13}$CO/\co\ ratio is about a factor 2-3, with the typical parameters of the region ($A_V\sim5-10$, $G_0\sim\pot{4}-\pot{5}$ and $n_\mathrm{H_2}\sim\pot{4}-\pot{5}$).
The results reveal that selective photodissociation does indeed occur south of the ionization front, and there is an indication that selective photodissociation is also taking place south of the elephant trunk.

At the position of the elephant trunk, at the IF and for clump E the ratio is smaller than the standard value. In this case, the ratio may still be influenced by optical depth effects (in particular an underestimated opacity for $^{13}$CO, which was determined from Eq.~\ref{eq:deteq}, assuming optically thin emission). 

\subsection{Ionization front and geometry of the region}\label{ssec:geomif}

Neither the present study nor that carried out by \citet{Massietal97} found significant quantities of molecular material around the ``Bar'' (see Fig.~\ref{fig:kbandif} and \ref{fig:cs-c18o}). This rules out the possibility that the ``Bar'' is an ionization front eroding a molecular cloud.
This also implies that we cannot exclude an association of G353.2+0.9 with Pis-24 based on the position of this feature (as argued by \citealt{Fellietal90} prior to the availability of molecular line maps of the region). Energetic, spectral, and excitation analyses \citep[e.g.][]{Massietal97, Bohigasetal04} indicate that Pis-24 is indeed associated with G353.2+0.9 and that their proximity is not just a projection effect.
The \textit{actual} ionization front in G353.2+0.9 lies along the IR-bright feature labeled IF in Fig.~\ref{fig:kbandif} and is possibly associated with the UC\hii~region C \citep{Fellietal90}.

We found a significant number of molecular cores along this bright ridge of emission and its continuation to the north-west, where the NIR brightness strongly diminishes. Furthermore, densities and temperatures are on average higher in this region.
Our application of RADEX also revealed slight increases in $\tk$, volume density of H$_2$, and opacity here. There is intense radio continuum emission at $5\usk\giga\hertz$ \citep{Fellietal90} associated with this feature. This, together with the molecular gas distribution and physical conditions, confirm that IF is the main ionization front in G353.2+0.9.
Here, the ionizing flux generated by the stars embedded in the region and by those of Pis-24,  erodes the molecular cloud and pushes its material towards the north. 
When the resolution is high enough, one can see that the clumps near the ionization front, such as H, D, and C, are roughly parallel to it. High-density tracers are observed in rather small features at the edge of the ionization front, indicating the regions where the gas was compressed by the shock front.

Molecular emission strongly decreases south of this ionization front and no massive clumps are observed here, although some emission is visible, especially in low-density tracers (see Fig.~\ref{fig:cs-c18o}). 
\citet{Bohigasetal04} found that the region immediately surrounding the ionization front is characterized by a thin layer ($\pot{-3} \usk \pc$) of very dense ionized material, where a photoevaporative flow is generated. 

We detect only very faint emission from CN(2-1) along the ``Bar'', which implies that there are very small clumps immersed in the PDR.

\subsection{Pismis-24 13 (N36)}

This star is located in the northern part of G353.2+0.9 (cf. Fig.~\ref{fig:kbandif}) and is classified as a spectral type O6.5 V((f)) \citep{Masseyetal01}. Pismis-24 13 is worth noting because it seems to have produced its own \hii\ region in the molecular gas (see e.g., Fig.~3 in \citealt{HesterDesch05}). This is confirmed by the radio continuum and ion-line observations \citep{Bohigasetal04}, which reveal free-free emission following very well the outer edge of the cavity and an increase in electron density in coincidence with this feature.
This \hii\ region appears to be in the foreground with respect to the elephant trunk and ionization front. 

Dense star clusters often produce runaway OB stars with high radial velocities through two different mechanisms \citep{ZinneckerYorke07}: asymmetric supernova explosions and dynamical three-body encounters. The radial velocities of these objects exceed $40\usk\kilo\metre\per\second$. \citet{Gvaramadzeetal11} confirm that NGC 6357 is rich in OB runaway stars ejected from the clusters within the cavity. Pismis-24 13 could be one of these OB runaways, ejected from Pis-24 in the direction of the molecular clouds, where its radiation and wind then created the observed \hii\ region. To confirm this hypothesis we need a spectral measurement of the radial velocity of the star.

\section{Summary}

We have observed the Galactic \hii\ region G353.2+0.9 in several molecular lines, which has  allowed us to distinguish at least 14 clumps in its associated molecular cloud. We have determined temperatures, densities, and masses of each clump. We also identified the location of the real ionization front in G353.2+0.9. There is a tendency for the clumps near to the ionization front to have redder velocities than those further to the north. This is especially noticeable in low-density tracers (cf. Fig.~\ref{fig:cs-c18o}), and is caused by the expansion of the ionized, overpressurized gas pushing the molecular material away from the observer.

Excitation temperatures derived from the ratios of the line temperatures of the \co(1-0) to the (2-1) lines indicate that $\tex$ is in the range $15-25\usk\kelvin$, with the higher values being found along the IF.
The temperatures derived from \metac, which is an effective tracer of kinetic temperature, are found to lie in the range of $\sim22-45\usk \kelvin$.

Assuming LTE and a constant $\tex=20\kel$, we derived the molecular column densities of $^{13}$CO (from \citealt{Massietal97}), \co, and H$_2$CO, thus obtaining maps of molecular abundances, from their ratios with the H$_2$ column density, which we derived from the APEX $870\usk\mum$ image, assuming $\tk=30\kel$ and a gas-to-dust ratio $\gamma=100$ (Fig.~\ref{fig:ab-map}). The maps show similar features, with a decrease in the molecular abundance in the region of the elephant trunk and the IF with respect to other region of intense molecular emission [$\sim$(75\asec,75\asec), E; $\sim$(-100\asec,125\asec), D, B]. The molecular abundances derived in this way are uncertain by at least a factor of two, owing to the variations in $\tk$ across the region. Nevertheless, the region of lower molecular abundance outlines the IF, clearly showing the area where the influence of early-type stars is the strongest.

Column densities of molecular hydrogen derived from \co\ and H$_2$CO, under the assumption of LTE and with the abundances calculated as described above, range between $\pot{20}-\pot{23}\usk\cm^{-2}$. The visual extinctions are proportional to the H$_2$ column density: typical values are in the range $5-30\usk\mg$ depending on the transition used, while the maximum values, estimated from H$_2$CO assuming LTE, are found in clumps C and E ($\sim 50\usk\mg$). The volume densities, determined assuming spherical geometry for the clumps, lie between $\sim\pot{3}\usk\cm^{-3}$ and $\sim\pot{5}\usk\cm^{-3}$. The volume density derived in this way is of the same order of magnitude as the critical density of the transition used to determine it.
 
The total mass of gas in the region is $\sim2000\usk \msun$. Single clumps have masses in the range $10-\mathrm{several}\,\times \usk \pot{2}\usk\msun$. The uncertainty in the mass for a given clump is dominated by the different physical conditions probed by the different transitions.

A simple virial analysis shows that all the clumps with masses above $50\usk\msun$, in addition to two with lower masses ($M\sim 20\usk\msun$ G1 and O) have $\alpha\approx 1$, thus indicating that these clumps might be gravitationally bound.

We performed a non-LTE analysis with RADEX, considering the four transitions of the two CS isotopologues in one case and the two transitions of CN in the other. 

To analyze the model results, we used a Bayesian approach, evaluating the likelihood $P(D|m,\tk,N,n)$ based on Eqns.~(\ref{eq:likelihood}) and (\ref{eq:likelihood-cn}). We used a constant \textit{prior} for all the parameters for CS and a Gaussian prior for $\tk$ ($\mu=35\kel$, $\sigma=30\kel$) for CN, according to our knowledge from CO, \metac, and the results from CS, to reduce the degeneration on this parameter. 

For CS, we found $\tk\sim11-45\kel$, depending on the clump considered. The H$_2$ number density typically ranges from $\mathrm{several}\,\times\pot{4}\cm^{-3}$ to $\mathrm{few}\,\times\pot{5}\cm^{-3}$, but exceeds $\pot{6}\cm^{-3}$ for clump E, where $\tk\sim11\kel$. The CS column density lies between $\sim2\times\pot{13}\cm^{-2}$ and $\sim2\times\pot{14}\cm^{-2}$. Making use of this result, we determined the abundance of CS, taking the ratio of the molecular column density to that of H$_2$ derived from the $870\mum$ emission. The abundances were found to lie between $7.0\times\pot{-10}$ and $4.0\times\pot{-8}$ (Table~\ref{tab:radex-cs}). These are lower limits  because we do not know the column density of each velocity component, but just that of the strongest one. However, the results should not change by much, given that at a certain offset there is usually a single velocity component that dominates the emission. 

For CN, we were able to construct maps for $\tk$, $N$, and $n$. We found that $\tk$ lies typically in the range of $25-32\kel$, with a maximum of $\sim45\kel$, but $\tk$ is the most poorly constrained parameter, thus making the maps quite difficult to interpret. The H$_2$ number densities that we found are on average higher than those suggested by CS, in the range of $\sim1-6\,\times\pot{5}\cm^{-3}$, which is consistent with the idea that most of the emission comes from high-density surface layers of the PDR \citep{Simon97}. The temperature $\tk$ shows a similar behaviour:  the temperatures found from CN are usually slightly higher than those derived from CS or \co. In the region of the elephant trunk and the IF, the volume density is even higher, $\sim\mathrm{few}\times\pot{6}\cm^{-3}$, which is similar to the values also reached for E and D. The column density lies between $\sim 5.0 \times \pot{13}\cm^{-2}$ and $\sim 3.0 \times \pot{14}\cm^{-2}$. Having a map of CN column density for each clump, we summed them at every position, obtaining an integrated $N_\mathrm{CN}$ map, similar to those of \co, $^{13}$CO, and H$_2$CO. This allowed us to obtain an abundance map (Fig.~\ref{fig:cn-ab-map}) that resembles those of the other molecules. The abundances found are in the range $5.0-7.9\times\pot{-9}$ (Table~\ref{tab:cn-radex}). This map showed that CN is enhanced in the region of the trunk and along the IF, where the decrease in molecular abundance is less pronounced than for other molecules.

We did not find significant quantities of molecular material in the region near the ``Bar'', which had been previously thought to be an ionization front. 
This is consistent with the idea that the ``Bar'' appears  to be a layer of ionized matter seen edge-on \citep{Bohigasetal04}, the result of the free wind from the massive stars of Pis-24 interacting with the photoevaporative flow generated at the true ionization front (see Fig.~\ref{fig:kbandif}). 
The presence of a very faint feature in CN(2-1) along the ``Bar'' possibly suggests that the small quantity of molecular material in this region could be distributed in very small condensations inside the PDR.

The high incident FUV flux strongly influences the shape and the properties of G353.2+0.9. We investigated the presence of selective photodissociation of \co\ making use of $^{13}$CO data of \citet{Massietal97}. The $^{13}$CO$/$\co\ ratio is nearly everywhere consistent with the standard value of the relative abundance of the two isotopologues. We found that south of the ionization front and the elephant trunk, \co\ is underabundant with respect to $^{13}$CO, being photodissociated by the strong, energetic radiation from Pis-24 stars.

There seems to be a separate semispherical \hii\ region in the northern part of G353.2+0.9, associated with the star Pis-24 13 (N36), which is an O6.5 V((f)) \citep{Masseyetal01}. We propose that this star is a runaway O star, dissociating the molecular gas while making its way through it. Radial velocity measurements are needed to confirm this hypothesis.

\begin{acknowledgements}
We thank the ATLASGAL project for kindly providing the APEX data. This research made use of the NASA ADS, SIMBAD and CDS (Strasbourg) databases. At the time the observations discussed here were performed, Achim Tieftrunk was employed by the I. Physikalische Institut in K\"oln. We also thank Malcolm Walmsley and the anonymous referee for their useful comments, that improved the initial paper.
\end{acknowledgements}

\bibliographystyle{aa.bst}
\bibliography{AGiannetti-G353.bib}

\end{document}